\documentclass{aa}
\usepackage{txfonts}
\usepackage{graphics}
\usepackage{graphicx}
\usepackage{amssymb}
\usepackage{afterpage}
\usepackage{natbib}
\begin{document}

\def\deg{$^{\rm o}$}
\def\ico{I$_{\rm CO}$}
\def\i14{I$_{\rm 1.4}$}
\def\Qco{Q$_{\rm CO/RC}$}
\def\qco{q$_{\rm CO/RC}$}
\def\qfir{q$_{\rm FIR/RC}$}
\def\matteo{Paper I}
\def\P{$\bar{\rm P}$}
\def\n0{N$_{0}$}
\def\tc{$t_{\rm c}$}
\def\tsyn{$t_{\rm syn}$}

\defcitealias{matteo05}{Paper~I}

\title{Thermal and non-thermal components 
of the interstellar medium at sub-kiloparsec scales in galaxies}

\author{ R. Paladino \inst{1,2} \and M. Murgia \inst{1,3} \and T.T. Helfer\inst{4} 
\and T. Wong \inst{5,6}\and R. Ekers\inst{5} \and L. Blitz\inst{4} \and L. Gregorini \inst{3,7}
\and L. Moscadelli\inst{1} 
}
\offprints{R. Paladino, rpaladin@ca.astro.it}

\institute{
INAF\,-\,Osservatorio Astronomico di Cagliari, Loc. Poggio dei Pini, Strada 54,
I-09012 Capoterra (CA), Italy
\and
Dipartimento di Fisica, Universit{\`a} di Cagliari, Cittadella Universitaria, 
I-09042 Monserrato (CA), Italy
\and
Istituto di Radioastronomia - INAF, Via Gobetti 101, I-40129 Bologna, Italy
\and
Radio Astronomy Laboratory, University of California, Berkeley, CA 94720, USA
\and
Australia Telescope National Facility, CSIRO, P.O. Box 76, Epping, N.S.W., 1710, 
Australia
\and
School of Physics, University of New South Wales, Sydney, NSW 2052,
Australia
\and
Dipartimento di Fisica, Universit{\`a} di Bologna, Via Irnerio, 46, 
I-40126 Bologna, Italy
}

\date{Received; Accepted}

\abstract{}{ We present new radio continuum observations of ten  BIMA SONG galaxies,
 taken at 1.4 GHz with the Very 
Large Array. These observations allow us to 
extend the study of the relationships between the  radio continuum (RC) and 
CO emission to 22 CO luminous galaxies for which 
single dish CO images have been added to interferometric data.  
New \emph{Spitzer} infrared (IR) images of six of these 
galaxies  have been released. 
The analysis of these high resolution images allowed us to probe the RC-IR-CO 
correlations down to linear scales of a few hundred pc.}
{We compare the point-by-point RC, CO and mid-IR intensities across entire galaxy
 disks, producing radial profiles and spatially resolved images of the RC/CO and 
RC/mid-IR ratios.}
{For the 22 galaxies analysed, the RC-CO correlation on  
scales from $\sim10$ kpc down to $\sim100$ pc is nearly linear 
and has a scatter of a factor of two, i.e. comparable to that of the global 
correlations. There is no evidence for any severe degradation of the 
scatter below the kpc scale. This also applies to the six galaxies for which 
high-resolution mid-IR data are
 available. In the case of \object{\object{NGC 5194}}, we find that the non-thermal radio spectral index 
 is correlated with the RC/FIR ratio. } 
{The scatter of the point-by-point correlations does not increase significantly 
with spatial resolution. We thus conclude that we have not yet probed the physical 
scales at which the correlations break down. However, we observe local  
deviations from the correlations in regions with a  high star formation rate, 
such as the spiral arms, where  we
 observe a flat radio spectrum and a low RC/FIR ratio. 
In the intra-arm regions and in the peripheral regions 
of the  disk, the RC/FIR is generally higher and it is 
characterized by a steepening of the radio spectrum.}

\maketitle

\section{Introduction}

One of the major goals in studying spiral galaxies is to understand the 
relationship between the star formation process and the physical conditions in the 
interstellar medium. Since the discovery that stars form in molecular clouds,
several efforts have been directed towards the study of the relation between
the emission of the CO molecule, which traces the bulk of molecular gas,
 and the other star formation indicators.

Global studies  have revealed relationships between the radio 
continuum (RC), far-infrared (FIR) and CO emissions.
 The global correlation between the FIR and centimeter-wavelength RC emission is 
one of the strongest correlations in extragalactic research. 
Early works on this topic were reported by \cite{helou85} and 
\cite{dejong85}.
\cite{condon92} and more 
recently \cite{yun01} found that these two emissions are linearly 
correlated over five orders of magnitude in luminosity, with an rms scatter of less 
than a factor of 2.
On global scales the FIR emission is well correlated with the CO emission 
\citep[e.g.][]{DevYou90}  and the CO emission is well correlated with the RC 
\citep{rickard77,israel84,adler91}.

What is most extraordinary about the FIR-RC or CO-RC correlations is that they couple 
emissions arising from completely different processes. The usual explanation of the FIR-RC 
 correlation invokes massive-star formation which accounts for the thermal radio 
emission via ionizing stars, for the non-thermal radio luminosity via 
supernova events, and for the
FIR luminosity by means of massive stars heating the dust \citep{wunderlich88}.
However, this basic scenario has too many steps and too many parameters to explain
 the tightness of the observed correlations.
More complex models have been put forward, see \citet[ hereafter Paper I]{matteo05}
for a brief summary, but a '{\it{definitive}}' model 
has not yet emerged.

Although there is still no consensus  on the causes of the RC-FIR-CO correlations,  
nor any adequate explanation for their tightness, these correlations have been used 
to determine empirically   the star formation rate (SFR) \citep{cram98} 
and the star formation efficiency (SFE)  in galaxies 
\citep{gao01,gao03,murgia02,rownd99}. 
 More generally it is found that
 these correlations  are well described by a ``Schmidt Law'' of 
the type SFR $ \propto \rho_{gas}^{N}$, where  $\rho_{gas}$ 
 is  the total gas density, and the exponent $N$ typically ranges from 1.3 
to 1.5 \citep{kennicutt98b}.

Detailed studies of individual objects at high spatial resolution are essential to 
understanding whether there is a common
 physical process behind the observed correlations and 
possibly synthesizing the various
empirical descriptions.

An improved approach to the study of the spatial properties of the FIR-RC 
correlation was presented in \cite{marsh95}. They studied the FIR-RC
correlation on intermediate scales of $\sim$ 1--3 kpc  and they found a small but
systematic decrease in the FIR-RC ratio as a function of increasing radial distance
from the center.
In contrast, \cite{murgia02} found that the ratio between the RC and CO emission
is constant, to within a factor of 3, both inside the same galaxy and from galaxy to
galaxy, down to kpc size scales.
\cite{leroy05} extended this result to a sample of dwarf galaxies, 
finding that this relationship between CO and RC is the same in 
all galaxies. In this context, dwarf galaxies appear to be simple low-mass versions 
of large spirals.

So far the low resolution of FIR observations has prevented  the comparison of
the FIR and RC emission  on kpc scales for all but a few Local Group sources.
\cite{xu92}, studying the Large Magellanic Cloud 
found evidence for a breakdown in the FIR-RC correlation 
at a linear scale of  $\sim$ 70 pc. 
In the Milky Way, \cite{boulanger88} analysed the Orion region and showed
that the FIR-RC correlation breaks down on the scale of a few hundred parsecs around
a region of star formation.

The situation has recently changed with 
 the advent of the satellite  {\it Spitzer} 
that provides  the necessary resolution of IR observations
 to extend the study of the correlation to sub-kpc scales in many 
external galaxies.  
 These observations
 can shed new light on the study of  the correlation on local scales.
Recently, \cite{murphy06} presented an initial look
at the far-infrared-radio correlation within the star-forming disks of four nearby 
galaxies using {\it Spitzer} images.

An ideal sample to investigate the spatially resolved FIR-RC-CO 
correlations is the BIMA SONG sample.
The Berkeley-Illinois-Maryland Association Survey of Nearby Galaxies (BIMA SONG)
is an imaging survey of  the CO (J=1-0) emission in 44 nearby spiral galaxies with a
typical angular resolution of 6\arcsec~\citep{regan01, helfer03}.
For half of the galaxies
in the BIMA sample single-dish CO images at 55$''$ resolution obtained
 with the NRAO 12 m telescope\footnote{The National Radio Astronomy
Observatory is a facility of the National Science Foundation operated
under cooperative agreement by Associated Universities, Inc.} have been added to
interferometric data. The combined images represent a significant improvement with
respect to the previous CO extragalactic surveys, since in each galaxy  a large area can
be mapped at high resolutions avoiding the missing flux problem.

Recently we  extended the study of CO-RC correlation
down to  a spatial resolution of $\sim$ 100 pc,
 by collecting 20 cm D and B-array data of a subsample of  nine bright BIMA
SONG galaxies available in the VLA archive (Paper I). For these nine galaxies it
has been possible
to compare  CO and RC images at two different angular resolutions: a low resolution
($\sim$ 55$''$) set obtained from the NRAO 12m and VLA D-array, and a high resolution
($\sim$ 6$''$) set.  The point-by-point RC and CO brightnesses across the entire galaxy
disks have been compared and the ratio of the CO and RC flux has been defined as a way
to characterize the CO-RC correlation.

In this work we present the study of 13 more  BIMA SONG galaxies 
which allow us to complete the study, started with Paper I, of 22  CO-luminous BIMA galaxies.

In Section 2 we summarize the observations used in this paper and the data reduction
procedure. Section 3 describes briefly how we performed the data analysis.
The results are presented in Section 4 for the whole sample, and
 are discussed in Section 5. 
We analysed the  24 $\mu$m-RC correlation 
for six galaxies of the sample, for which high resolution
{\it Spitzer} mid-IR images are available, 
at a spatial resolution ranging from 
$\sim$400 pc to $\sim$100 pc. We present the 
results of this analysis in Section 6.
In Section 7 we compare the expectations of the leaky box diffusion model  
model, proposed in \matteo, with the observed trend of spectral index
 versus RC-FIR ratio for \object{NGC 5194}. 
In Section 8 we summarize our results. 

\begin{table*}[htbp]
\caption{Galaxy sample }
\begin{center}
\begin{tabular}{llllllll}
\\
\hline
\smallskip
Source  & $\alpha (J2000)$& $\delta (J2000)$ &d &i&$P.A.$& RC3 Type, Nucleus\\
\smallskip
name &  (h m s)&($^{\circ}$\,$'$\, $''$)& (Mpc)&$(^{\circ})$&$(^{\circ})$  &  & \\
\hline
\\
\object{NGC 0628}&01 36 41.7 & +15 46 59 &7.3 &24 & 25 &SA(s)c\\
\object{NGC 1068}&02 42 40.7 & - 00 00 48 &14.4& 33& 13&(R)SA(rs)b, Sy1.8\\
\object{NGC 2903}&09 32 10.1 & +21 30 02 &6.3 &61 & 17&SAB(rs)bc,HII\\
\object{NGC 3351}&10 43 31.1 & +11 42 14 &7.4 &40 & 13&SB(r)b,HII\\
\object{NGC 3521}&11 05 48.0 &- 00 02 04 &7.2 &58 & 164&SAB(rs)bc,HII/L2\\
\object{NGC 3627}&11 20 15.1 & +12 59 22 &11.1 &63 & 176& SAB(s)b, T2/Sy2\\
\object{NGC 3938}&11 52 49.6 & +44 07 14 &17 &24 & 0& SA(s)c,HII\\
\object{NGC 4303}& 12 21 54.9 & +04 28 25 & 15.2 & 27 & 0 & SAB(rs)bc, T2\\
\object{NGC 4321}&12 22 54.8 & +15 49 20 &16.1 &32 & 154& SAB(s)bc,T2\\
\object{NGC 4826}&12 56 43.6 & +21 40 59 &4.1 &54 & 111&(R)SA(rs)ab,T2\\
\object{NGC 5248}&13 37 32.06 & +08 53 07 &22.7 &43 & 110& SAB(rs)bc, HII\\
\object{NGC 5457}&14 03 12.48 & +54 20 55 &7.4 &27 & 40& SAB(rs)cd\\
\object{NGC 7331}&22 37 04.09 &+34 24 56 &15.1&62 & 172&SA(s)b,T2\\
\\
\hline
\\
\end{tabular}
\end{center}
\begin{list}{}{}
\item[]
Coordinates, distances, inclinations, position angles, RC3 type and nuclear
classifications from \cite{helfer03}. Inclination is defined assuming 
i=0\degr\ for face-on galaxies. Position angles are measured in 
degrees from north  through east. For nuclear spectral classification:
H II = H II nucleus, Sy = Seyfert nucleus, L = LINER, and T=transition object,
which has [O I] strengths intermediate between those of H II nuclei and LINERs
\end{list}
\label{Gala}
\end{table*}

\begin{table*}[htbp]
\caption{VLA B-array observation summary}
\begin{center}
\begin{tabular}{lllll}
\\
\hline
\smallskip
Source   & Date & Duration & Beam & $\sigma_{I}^{~a)}$\\
\smallskip
name &   & hours &HPBW($''$) & ($\mathrm{\mu Jy\;beam^{-1}}$)\\
\hline
\\
\object{NGC 0628} & 03-Nov.-2003 & 1.5&$5.7\times 5.2$& $~40$\\
\object{NGC 2903} & 04,06-Nov-2003 & 1.0&$5.9\times5.3$& $~35$ \\
\object{NGC 3351} & 28-Jul-1986$^{*}$&1.8&$4.5\times4.1$&  $~46$\\
\object{NGC 3521} & 04,06-Nov-2003 & 1.0&$4.9\times4.8$& $~65$\\
\object{NGC 3627} & 04,06-Nov-2003 & 1.0&$5.8\times5.7$& $~35$\\
\object{NGC 3938} & 04,06,10-Nov-2003 & 2.0 &$5.8\times4.9$& $~50$\\
\object{NGC 4303} & 13-Sep-1982$^{**}$ & 1.0 & $6.5\times4.6$&$~100$\\
\object{NGC 4321} & 04,06,10-Nov-2003 & 1.2 &$5.8\times5.1$& $~30$\\
\object{NGC 4826} & 06,10-Nov-2003 & 1.0&$5.5\times5.2$& $~45$\\
\object{NGC 5248} & 04,06,10-Nov-2003 & 1.8&$6.1\times5.2$& $~50$ \\
\object{NGC 5457} & 06-Aug-1990$^{***}$& 1.0&$5.8\times5.3$& $~50$\\
\object{NGC 7331} & 03-Nov.-2003 &1.0&$5.3\times4.8$& $~50 $\\

\\
\hline
\\
\end{tabular}\\
$^{*}$ Archival data: program A073;\\
$^{**}$ Archival data: program COND;\\
$^{***}$ Archival data: program AV180;
\end{center}
\begin{list}{}{}
\item[$a)$] Final noise of total intensity images.
\end{list}
\label{Obs}
\end{table*}

\begin{figure*}[t]
\begin{center}
\includegraphics[width=0.86\textwidth]{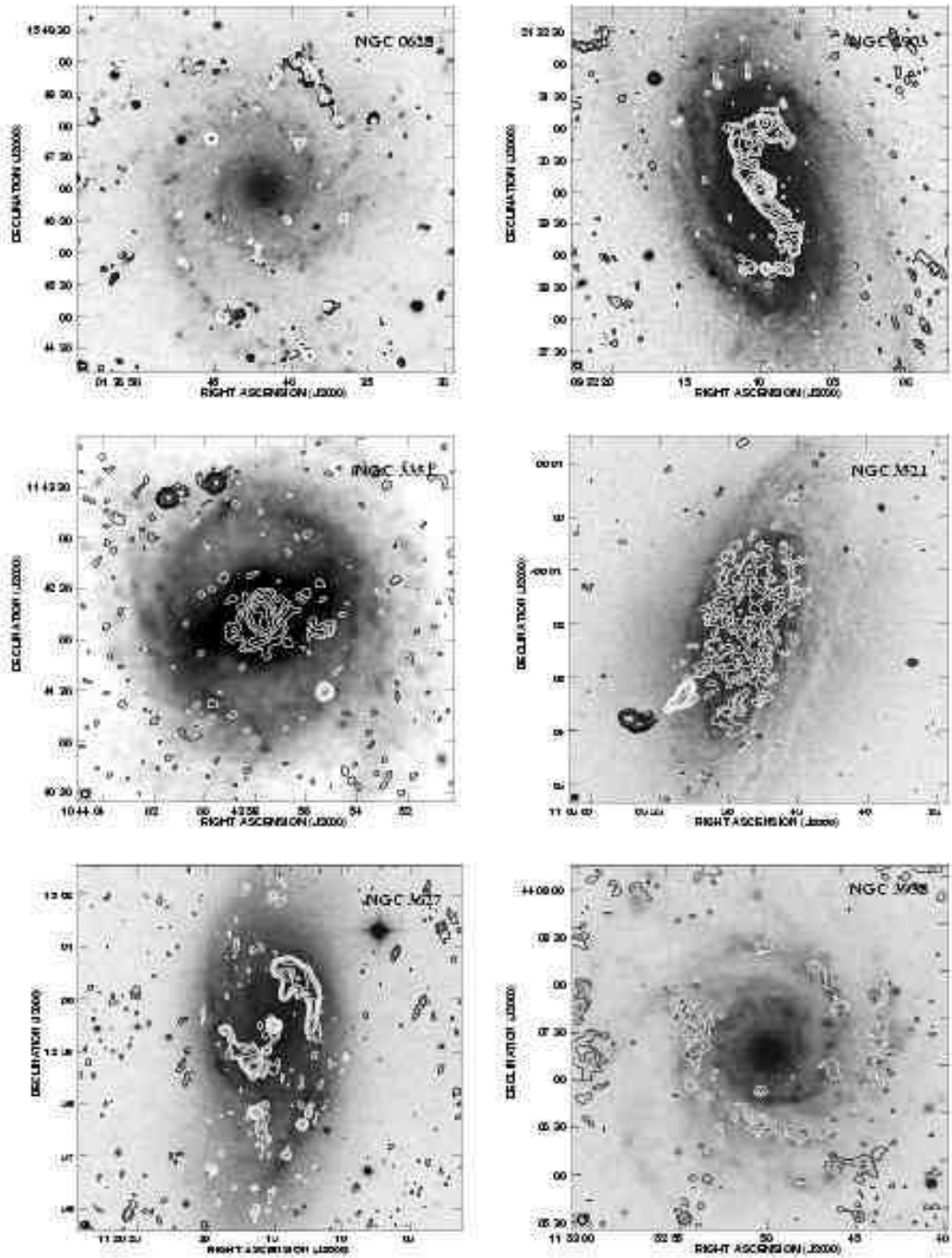}
\end{center}
\caption[]{$(a)$ Radio contours of galaxies overlaid on the optical images 
(grey scale),
taken from the Digitized Sky Surveys (DSS) archive.
See Table \ref{Obs} for sensitivity and beam sizes of radio images.  
 Contours start at 2~$\sigma$, and increase by steps of a factor of 2.}
\label{figura1a}
\end{figure*}
\setcounter{figure}{0}

\begin{figure*}[t]
\begin{center}
\includegraphics[width=0.86\textwidth]{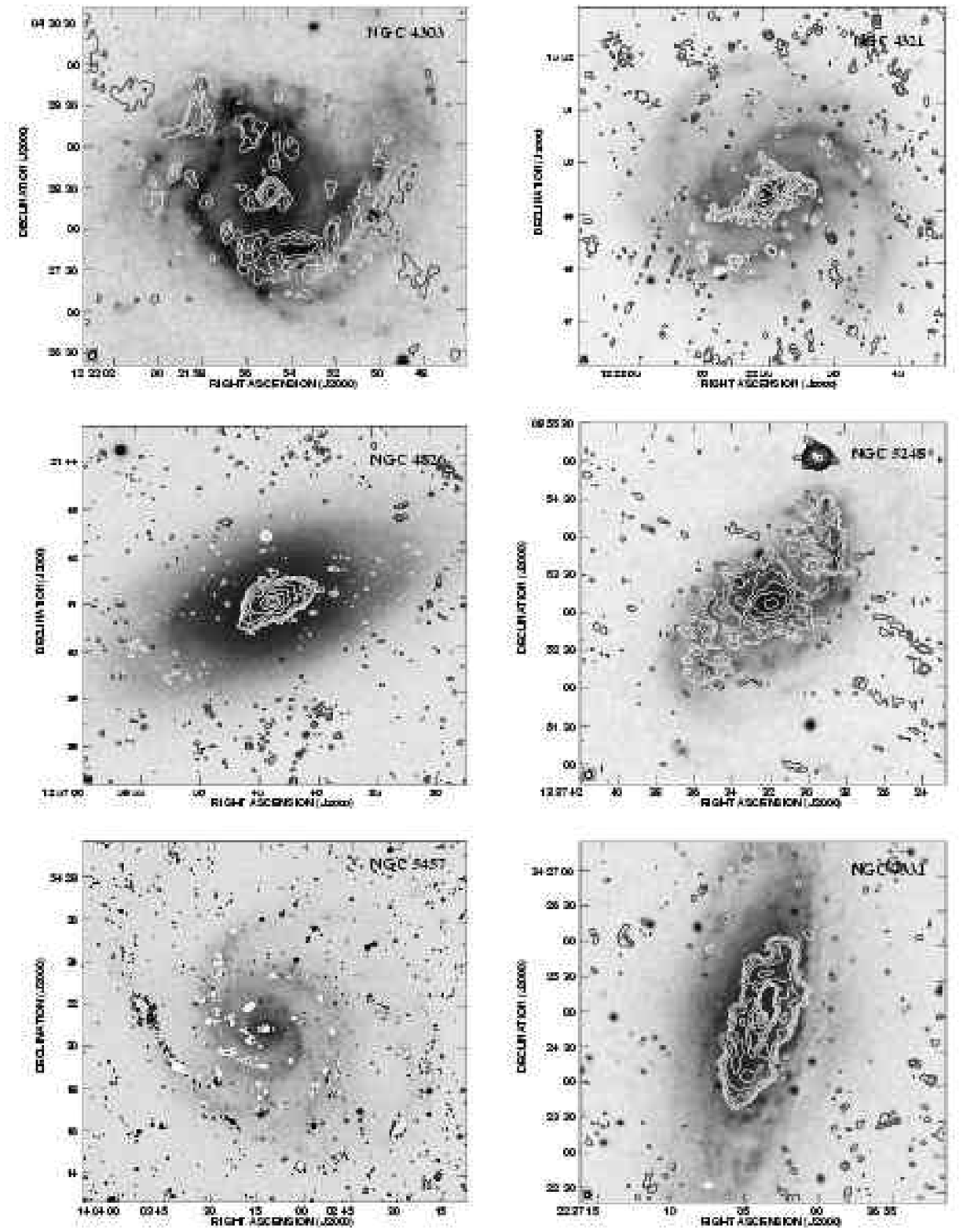}
\end{center}
\caption[]{$(b)$ See Fig. \ref{figura1a}.$a$. }
\end{figure*}

\begin{figure*}[ht!]
\begin{center}
\includegraphics[width=16.5cm]{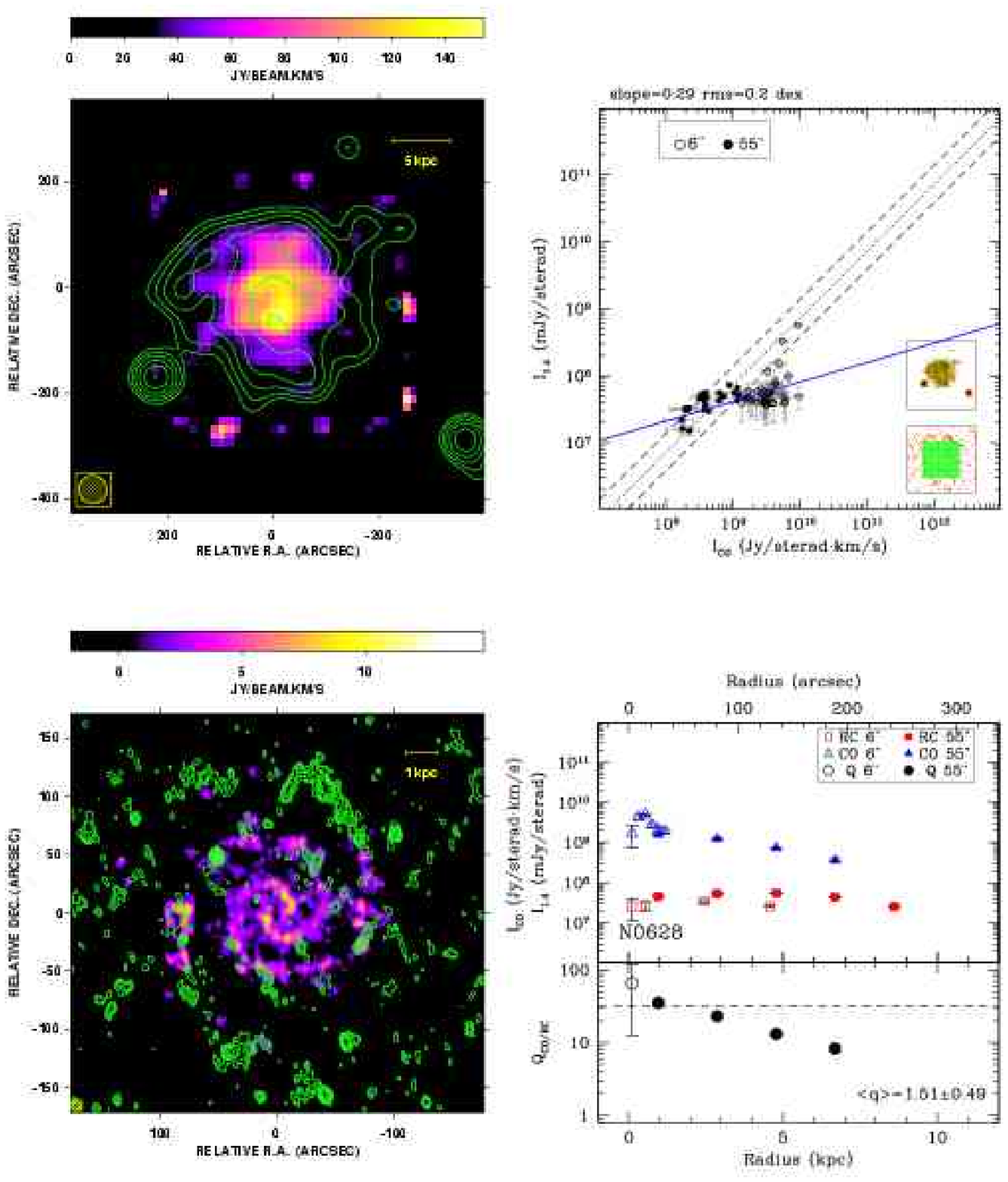}
\end{center}
\caption[] {
{\bf (a) \object{NGC 0628}.}  Radio continuum and CO comparison in BIMA SONG galaxies.
{\it (Upper left)} RC contours overlaid on CO emission for 55\arcsec\ images.
{\it (Lower left)} RC contours overlaid on CO emission for 6\arcsec\ images.
Contours start at 5$\sigma$ and scale by a factor of $\sqrt{2}$.
{\it (Upper right)} \i14\ as a function of \ico\  for the 55\arcsec\ and
6\arcsec\ data sets.  The insets show the box grids used; the grid size
corresponds to the resolution, so that each point is essentially
independent.
The dotted (dashed) lines show the mean (dispersion) of the
\i14\ -- \ico\ relation  obtained for the  sample containing 22 galaxies
(this work and \matteo).  The solid blue line is a weighted fit
to the points  which takes into account the errors in both coordinates.
{\it (Lower right)} Radial profiles of \ico\ and \i14\ {\it (top)} and
\Qco\ $\equiv$ \ico/\i14 {\it (bottom)} for the 55\arcsec\ and 6\arcsec\ images.
The mean and dispersion of \qco\ $\equiv \log (Q_{\rm CO/RC})$ is also given.
 }
\label{gal1}
\end{figure*}

\section{Observations and data reduction}

We present new radio observations of ten BIMA SONG galaxies, taken at 
1.4 GHz with the Very Large Array synthesis telescope in its B configuration.
These observations allow us to extend the study of the relationships between the radio 
continuum and CO emission, started in \matteo, to 22 CO-luminous galaxies, for 
which single dish CO images have been added to interferometric data.

To these newly observed we add three more BIMA SONG galaxies: 
NGC\, 3351, \object{NGC 4303} and NGC\, 5457
whose archival data were available, and \object{NGC 1068}, for which we use the high resolution
image of the FIRST public survey of the VLA.
The galaxies studied in this work are listed in Table \ref{Gala}, 
along  with their main geometrical parameters.
A summary of the VLA-B array data, including  date and length of observations, and beam
 size and rms level of produced images for each target is reported in Table \ref{Obs}.

We selected a bandwidth of 50\,MHz for the two IF channels centered at 1.465 and 1.385
 GHz.  The second of the two IFs was corrupted by interferences and had to be discarded,
reducing the signal to noise ratio
  by a factor of $\sqrt{2}$. Nevertheless, the time spent on each source
  was sufficient to guarantee that the dynamic  range of RC images is
 larger than that of the corresponding CO images.

The visibility data were calibrated, Fourier transformed, and deconvolved with the NRAO package AIPS following
 standard procedures.
Self-calibration was applied to
minimize the effects of phase variations.
The final cleaned maps were convolved
 with the beams listed in Table \ref{Obs}.
 The flux densities were brought to the scale of \cite{baars77}  using the flux
calibrators 3C48, 3C138 and 3C286.

In Fig. \ref{figura1a} we present the high-resolution radio 
image overlaid on the optical image for each galaxy, except for \object{NGC 1068},
for which we use the FIRST image.
We can recognize different morphologies of radio emission, which 
are described in the notes on individual galaxies reported in 
Section \ref{note}.

To improve the sensitivity of the radio continuum images to large-scale
structure, D and C array
 data were combined with the B array data.
When u-v data were available in the VLA archive we combined them in the visibility
 plane; otherwise D images were added in the image plane using the 
linear combination method implemented in the MIRIAD task IMMERGE 
\citep{miriad95}.

We convolved the original RC and CO maps to common beam sizes for all but
a few cases in which the data sets had intrinsic beam areas matching
within 10$\%$.
As in Paper I, we have CO and RC images at two different angular scales: low
resolution (~55$''$)  and high resolution 
(~6$''$).
The image parameters are listed in Tables \ref{tab_lr} and \ref{hr}, 
respectively.
The beam sizes are listed in angular and linear scales.

\begin{table*}[htbp]
\caption{Low resolution CO and RC image parameters.}
\begin{center}
\begin{tabular}{lllllll}
\noalign{\smallskip}
\hline                  
\noalign{\smallskip}    
Galaxy   & VLA data      &   RC-$\theta_{maj}\times\theta_{min}$  &  $\sigma_{RC}$  &
  CO-$\theta_{maj}\times\theta_{min}$ & $D_{maj}\times D_{min}^{~a)}$ & 
$\sigma_{CO}^{~b)}$      \\
         &               &   (\arcsec$\times$\arcsec)                 &     (mJy~bm$^{-1}$)     &    
(\arcsec$\times$\arcsec)    & (kpc$\times$kpc)    &   (Jy~bm$^{-1}$~km~s$^{-1}$)  
\\ \hline
\object{NGC 0628}  & D & 60$\times$60 & 0.30& 55$\times$55 & 2.0$\times$2.0 &  10.1\\  
\object{NGC 1068} & D & 60$\times$60 & 0.20 & 55$\times$55 & 3.8$\times$3.8 & 31.2\\
\object{NGC 2903} & D & 60$\times$60 & 0.20 & 55$\times$55 & 1.7$\times$1.7 &  36.1\\
\object{NGC 3351} & D & 54$\times$54 & 0.30 & 55$\times$55 & 1.9$\times$1.9 &  33.6\\
\object{NGC 3521} & D & 60$\times$60 & 0.15& 55$\times$55 &  1.9$\times$1.9 &  44.9\\
\object{NGC 3627} & D & 54$\times$54 & 0.20& 55$\times$55 & 2.9$\times$2.9 &  55.1\\
\object{NGC 3938} & D & 54$\times$54 & 0.15 & 55$\times$55 &4.5$\times$4.5 &  18.4\\
\object{NGC 4303} & D & 54$\times$54 & 0.40 & 55$\times$55 &4.0$\times$4.0 &19.6\\
\object{NGC 4321} & D & 54$\times$54 & 0.30 & 55$\times$55 & 4.3$\times$4.3 & 37.5\\
\object{NGC 4826} & D & 60$\times$60 & 0.10 & 55$\times$55 & 1.1$\times$1.1 &  26.5\\
\object{NGC 5248} & D & 54$\times$54 & 0.15 &55$\times$55 & 6.0$\times$6.0 &  28.3\\
\object{NGC 5457} &D & 48$\times$48 & 0.25 & 55$\times$55 &2.0$\times$2.0 &  24.7\\
\object{NGC 7331} & D& 48$\times$48 & 0.25 & 55$\times$55 & 4.0$\times$4.0 &  47.0\\
    
\hline
\noalign{\smallskip}
\label{tab_lr}
\end{tabular}

\end{center}
\begin{list}{}{}
\item[$a)$] The linear resolution refers to the CO beam.
\item[$b)$] $\sigma_{CO}$ is rms level in unclipped image of integrated intensity.
\end{list}
\end{table*}

\begin{table*}[tbp]
\caption{High resolution CO and RC image parameters.}
\begin{center}
\begin{tabular}{lllllll}
\noalign{\smallskip}
\hline                  
\noalign{\smallskip}    
Galaxy   & VLA data      &   RC-$\theta_{maj}\times\theta_{min}$  &  $\sigma_{RC}$  &  
CO-$\theta_{maj}\times\theta_{min}$ & $D_{maj}\times D_{min}^{~a)}$ & 
$\sigma_{CO}^{~b)}$      \\
         &               &   (\arcsec$\times$\arcsec)                 &     (mJy~bm$^{-1}$)     &    
(\arcsec$\times$\arcsec)    & (pc$\times$pc)    &   (Jy~bm$^{-1}$~km~s$^{-1}$)  
\\ \hline
\object{NGC 0628}  & B+D & 7.2$\times$5.3 & 0.035& 7.2$\times$5.3 & 250$\times$190&1.3  \\
\object{NGC 1068} & B & 8.8$\times$5.5 & 0.3 &8.8 $\times$5.5 & 614$\times$384 &5.3 \\  
\object{NGC 2903} & B+D & 6.8$\times$5.7 & 0.03 & 6.8$\times$5.5 & 210$\times$170&4.6  \\
\object{NGC 3351} & B+D & 7.3$\times$5.2 & 0.05 & 7.4$\times$5.2 & 265$\times$187&3.6  \\
\object{NGC 3521} & B+D & 8.8$\times$5.7 & 0.05 & 8.8$\times$5.7 & 310$\times$200&5.5  \\
\object{NGC 3627} & B+D & 7.3$\times$5.7 & 0.03 & 7.3$\times$5.8 & 390$\times$310 &2.8  \\
\object{NGC 3938} & B+D & 5.8$\times$4.9 & 0.03 & 5.9$\times$5.4 & 480$\times$440 &1.9  \\
\object{NGC 4303} & B+D & 7.3$\times$5.4 & 0.1  & 7.3$\times$5.4 & 538$\times$398&2.1\\
\object{NGC 4321} & B+D & 7.2$\times$5.3 & 0.03 & 7.2$\times$4.9 & 562$\times$382 &2.5 \\
\object{NGC 4826} & B+D & 7.5$\times$5.4 & 0.04 & 7.5$\times$5.4 & 150$\times$100 &4.9  \\
\object{NGC 5248} & B+D & 6.9$\times$5.7 & 0.06 & 6.9$\times$5.7 & 759$\times$627 &2.1  \\
\object{NGC 5457} & B+D & 5.8$\times$5.3 & 0.04 & 5.7$\times$5.4 & 210$\times$190 &1.6  \\
\object{NGC 7331} & B+C& 6.1$\times$4.9 & 0.03 & 6.1$\times$4.9 & 440$\times$360 &3.8  \\
    
\hline
\noalign{\smallskip}
\label{hr}
\end{tabular}

\end{center}
\begin{list}{}{}
\item[$a)$] The linear resolution refers to the CO beam.
\item[$b)$] $\sigma_{CO}$ is rms level in unclipped image of integrated intensity.
\end{list}
\end{table*}

\section{Data analysis }
\setcounter{figure}{1}
\begin{figure*}[]
\begin{center}
\includegraphics[width=16.5cm]{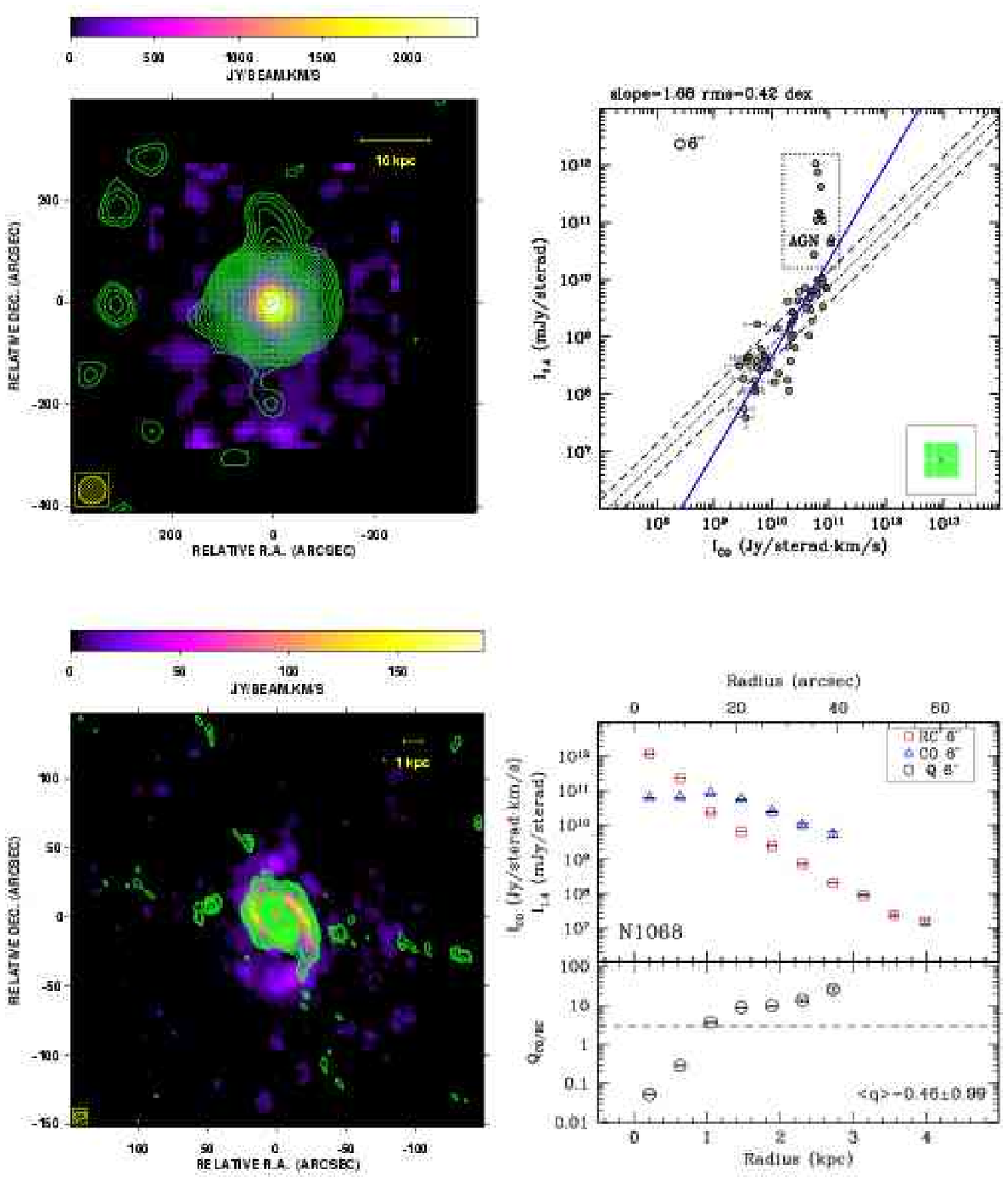}
\end{center}
\caption[] {{\bf (b) \object{NGC 1068}.} See Fig. \ref{gal1}$a$ for technical details.

}
\end{figure*}

\setcounter{figure}{1}
\begin{figure*}[]
\begin{center}
\includegraphics[width=16.5cm]{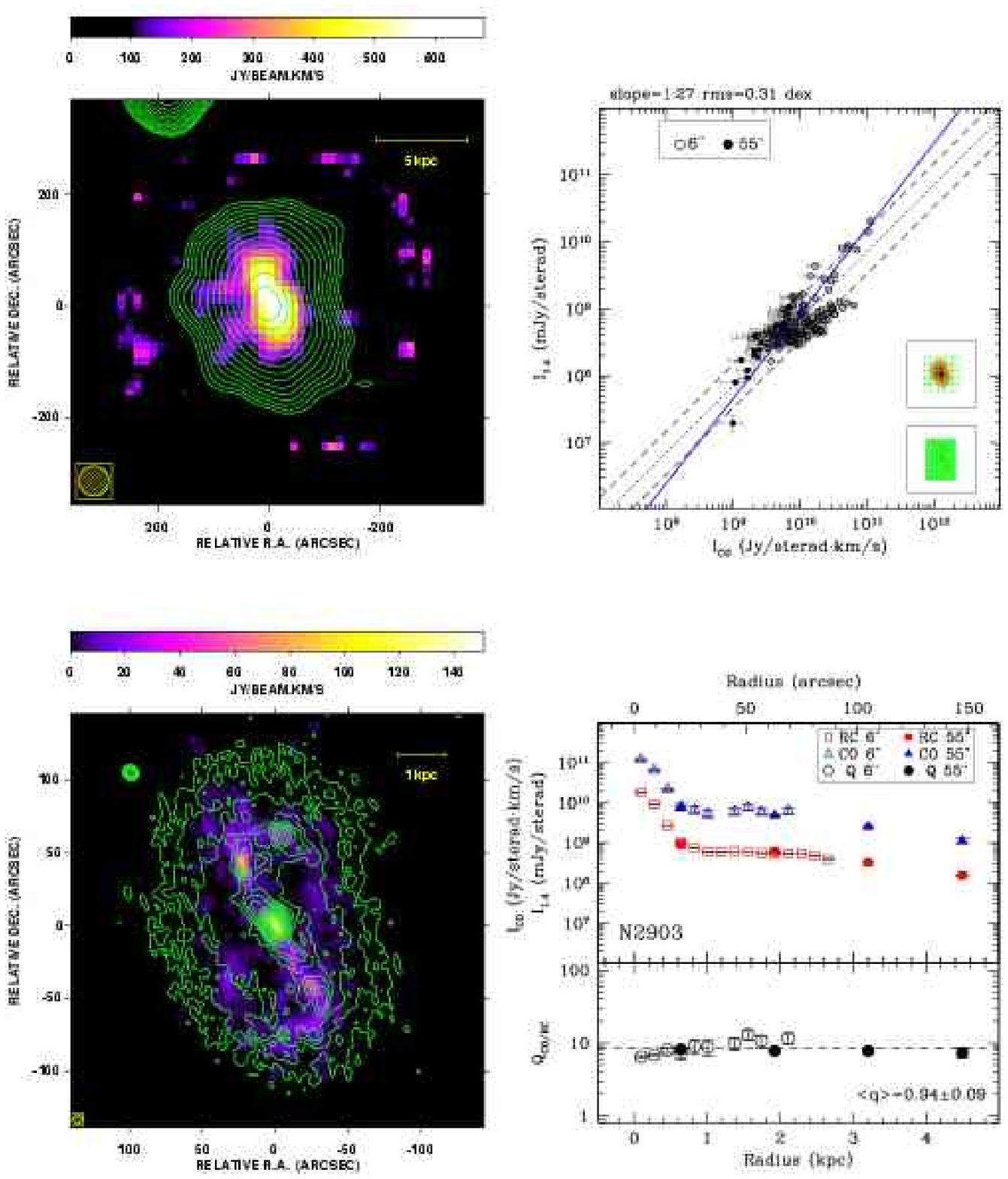}
\end{center}
\caption[] {{\bf (c) \object{NGC 2903}.} See Fig. \ref{gal1}$a$ for technical details.
 }
\end{figure*}

\setcounter{figure}{1}
\begin{figure*}[]
\begin{center}
\includegraphics[width=16.5cm]{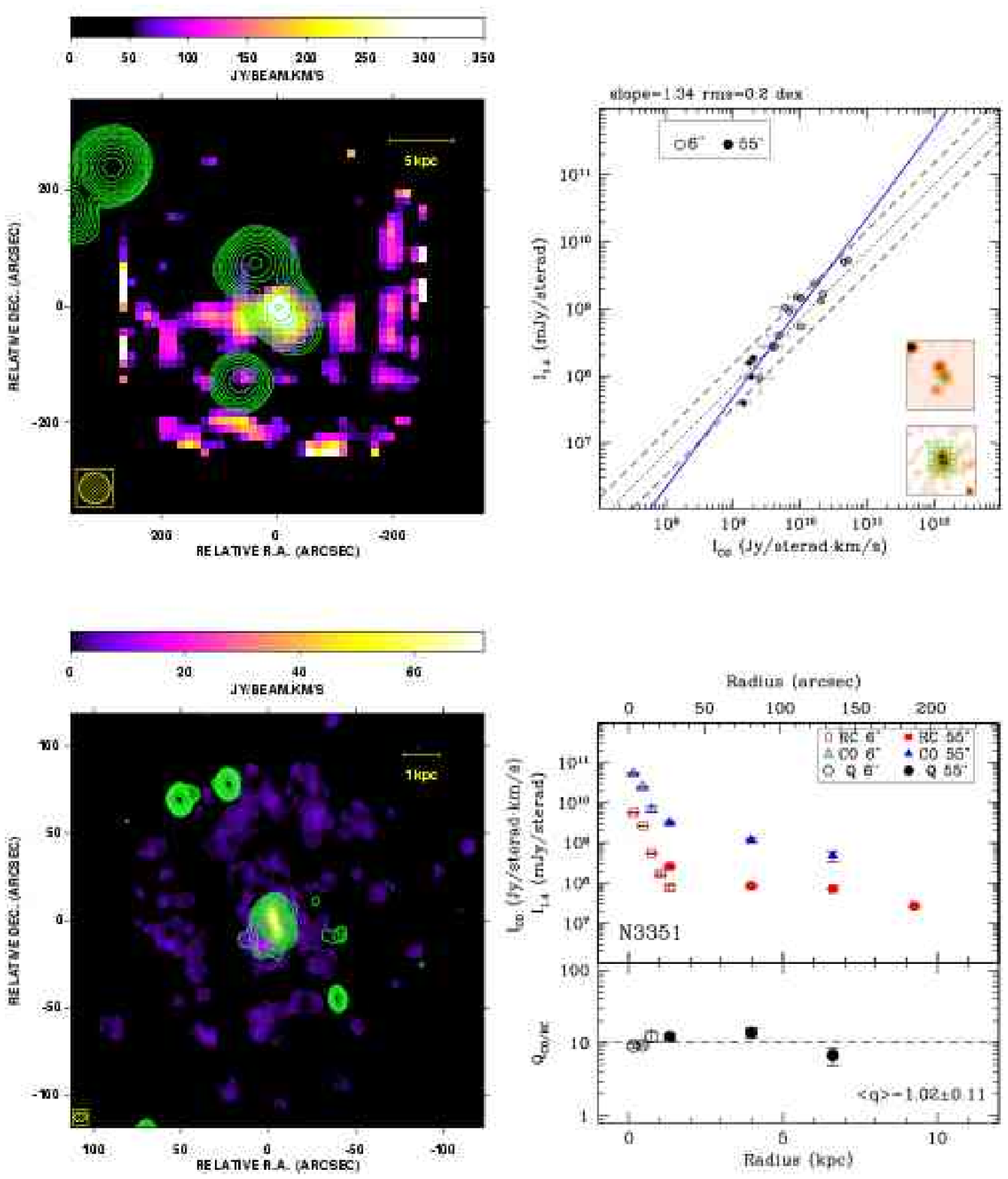}
\end{center}
\caption[] {{\bf (d) \object{NGC 3351}.} See Fig. \ref{gal1}$a$ for technical details.
}
\end{figure*}

\setcounter{figure}{1}
\begin{figure*}[]
\begin{center}
\includegraphics[width=16.5cm]{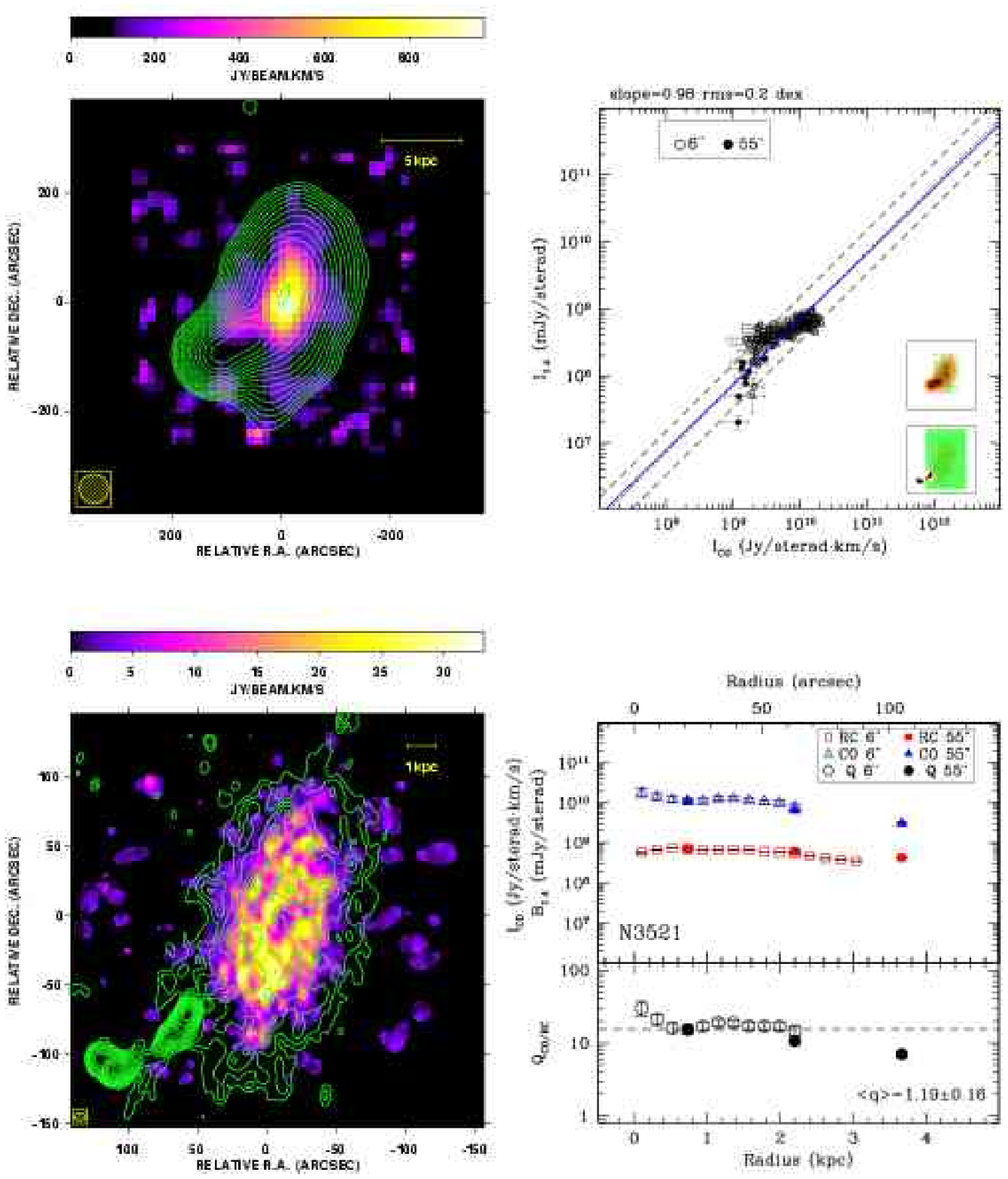}
\end{center}
\caption[] {{\bf (e) \object{NGC 3521}.} See Fig. \ref{gal1}$a$ for technical details.
}
\end{figure*}

\setcounter{figure}{1}
\begin{figure*}[]
\begin{center}
\includegraphics[width=16.5cm]{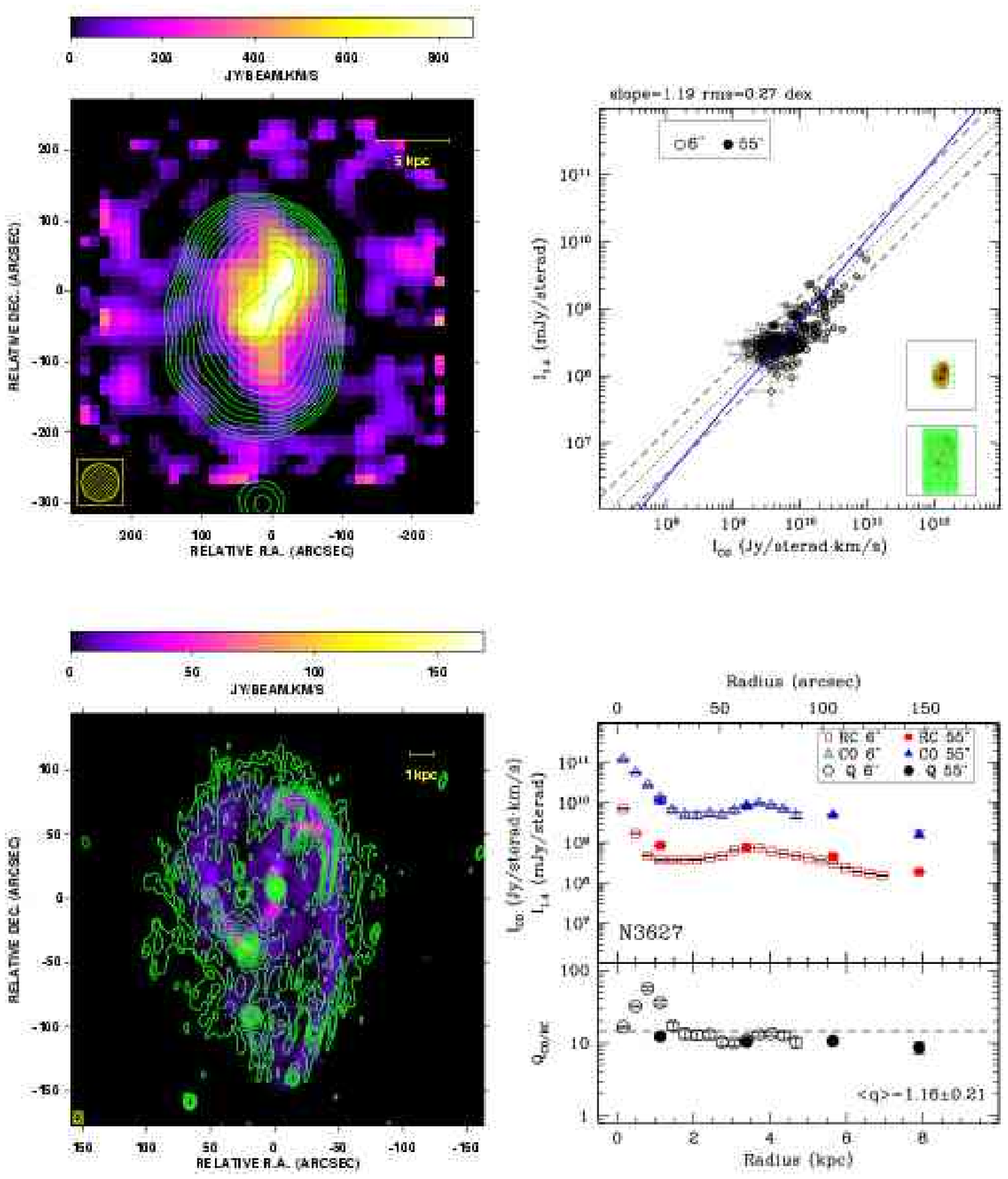}
\end{center}
\caption[] {{\bf (f) \object{NGC 3627}.} See Fig. \ref{gal1}$a$ for technical details.

}
\end{figure*}

\setcounter{figure}{1}
\begin{figure*}[]
\begin{center}
\includegraphics[width=16.5cm]{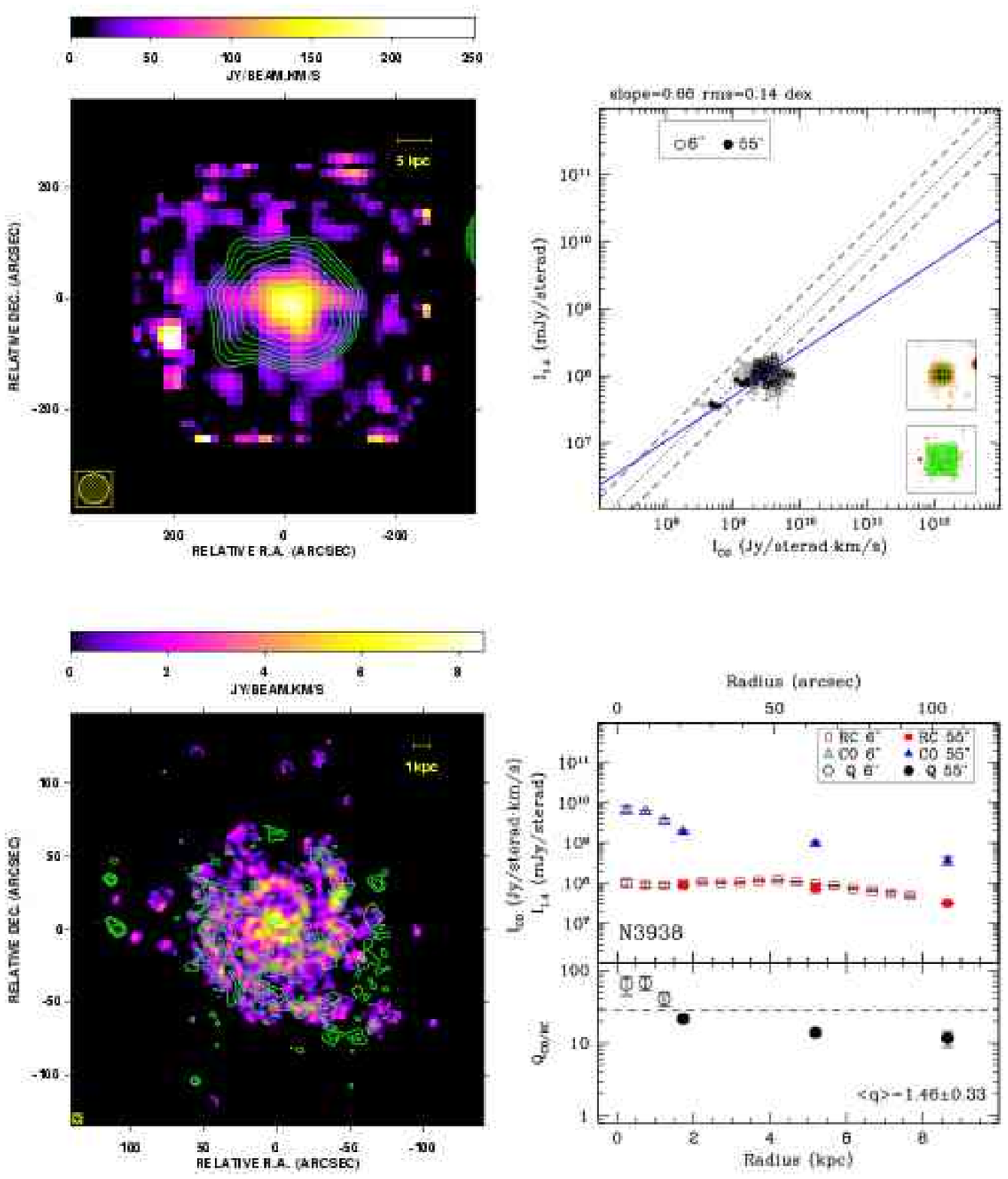}
\end{center}
\caption[] {{\bf (g) \object{NGC 3938}.} See Fig. \ref{gal1}$a$ for technical details. 
}
\end{figure*}

\setcounter{figure}{1}
\begin{figure*}[]
\begin{center}
\includegraphics[width=16.5cm]{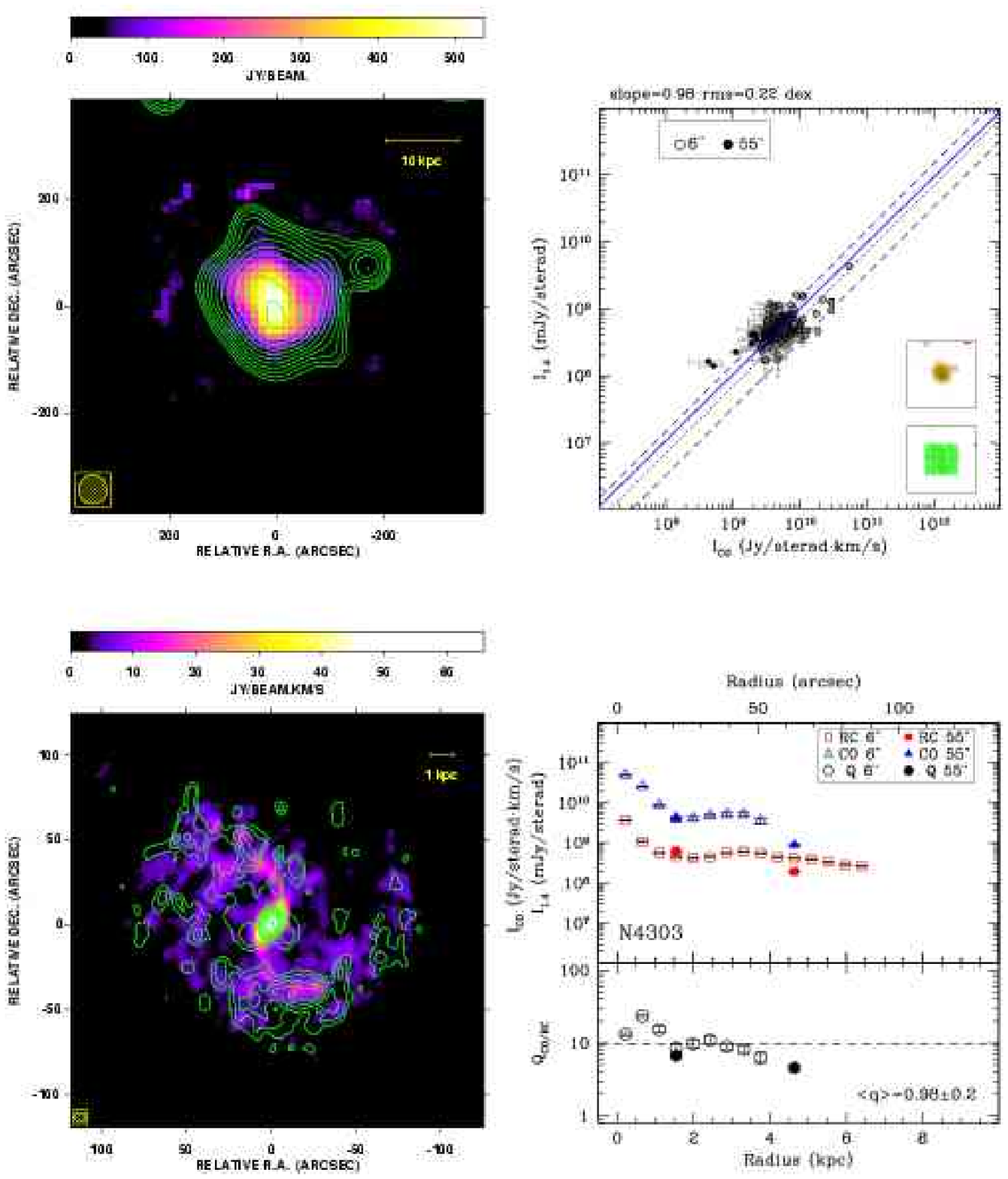}
\end{center}
\caption[] {{\bf (h) \object{NGC 4303}.} See Fig. \ref{gal1}$a$ for technical details.

}
\end{figure*}

\setcounter{figure}{1}
\begin{figure*}[]
\begin{center}
\includegraphics[width=16.5cm]{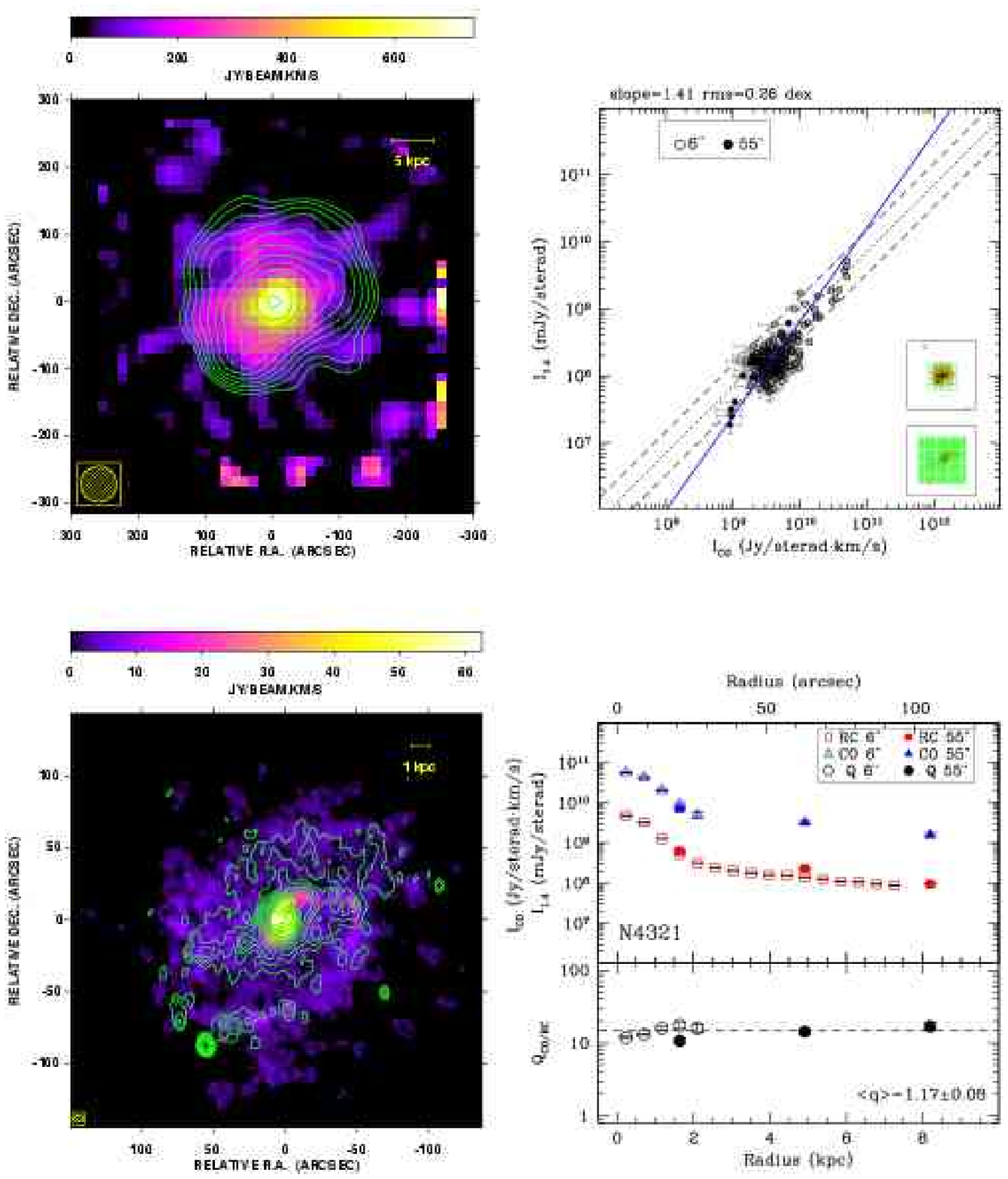}
\end{center}
\caption[] {{\bf (i) \object{NGC 4321}.} See Fig. \ref{gal1}$a$ for technical details.
}
\end{figure*}

\setcounter{figure}{1}
\begin{figure*}[]
\begin{center}
\includegraphics[width=16.5cm]{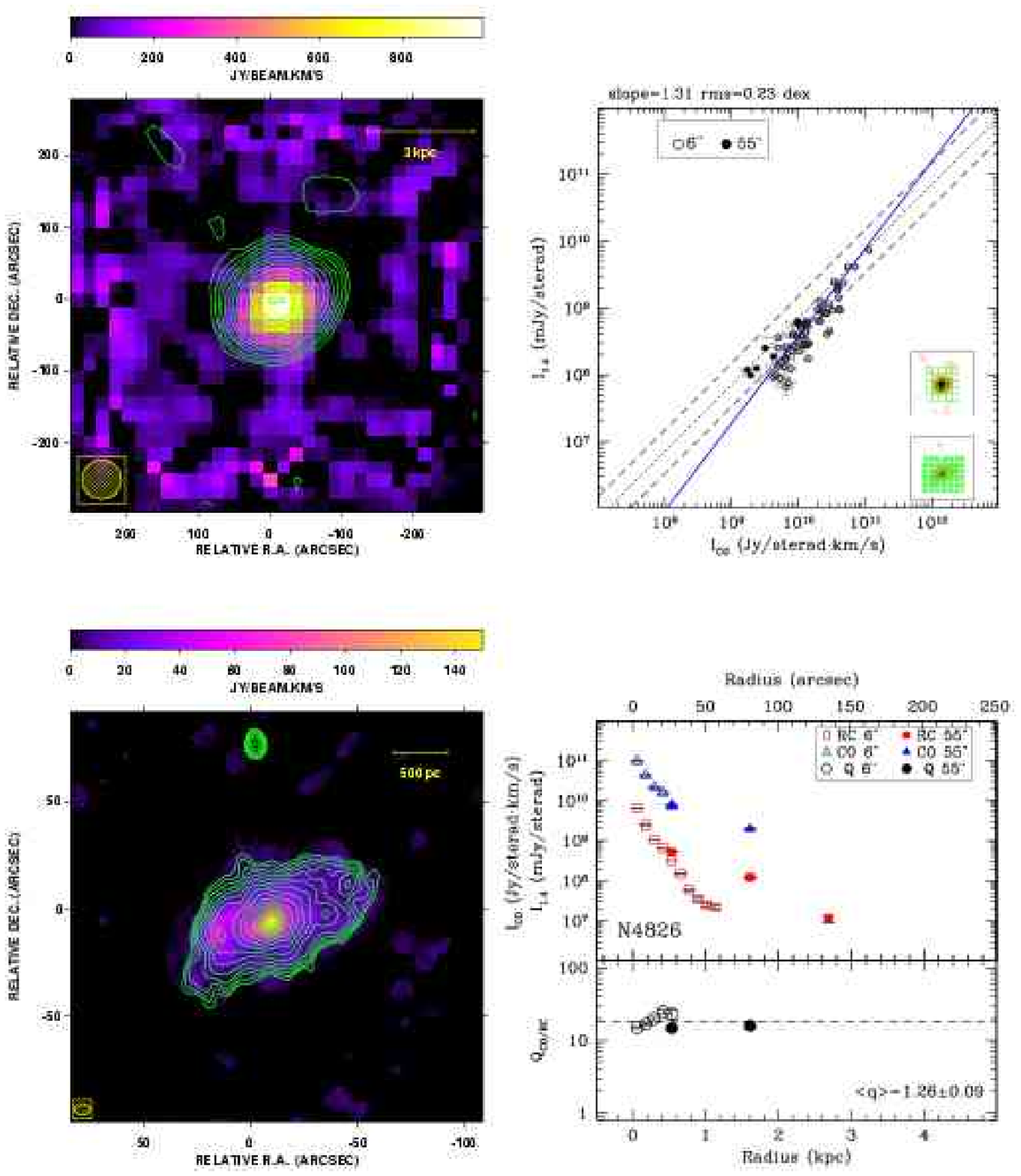}
\end{center}
\caption[] {{\bf (l) \object{NGC 4826}.} See Fig. \ref{gal1}$a$ for technical details.

}
\end{figure*}

\setcounter{figure}{1}
\begin{figure*}[]
\begin{center}
\includegraphics[width=16.5cm]{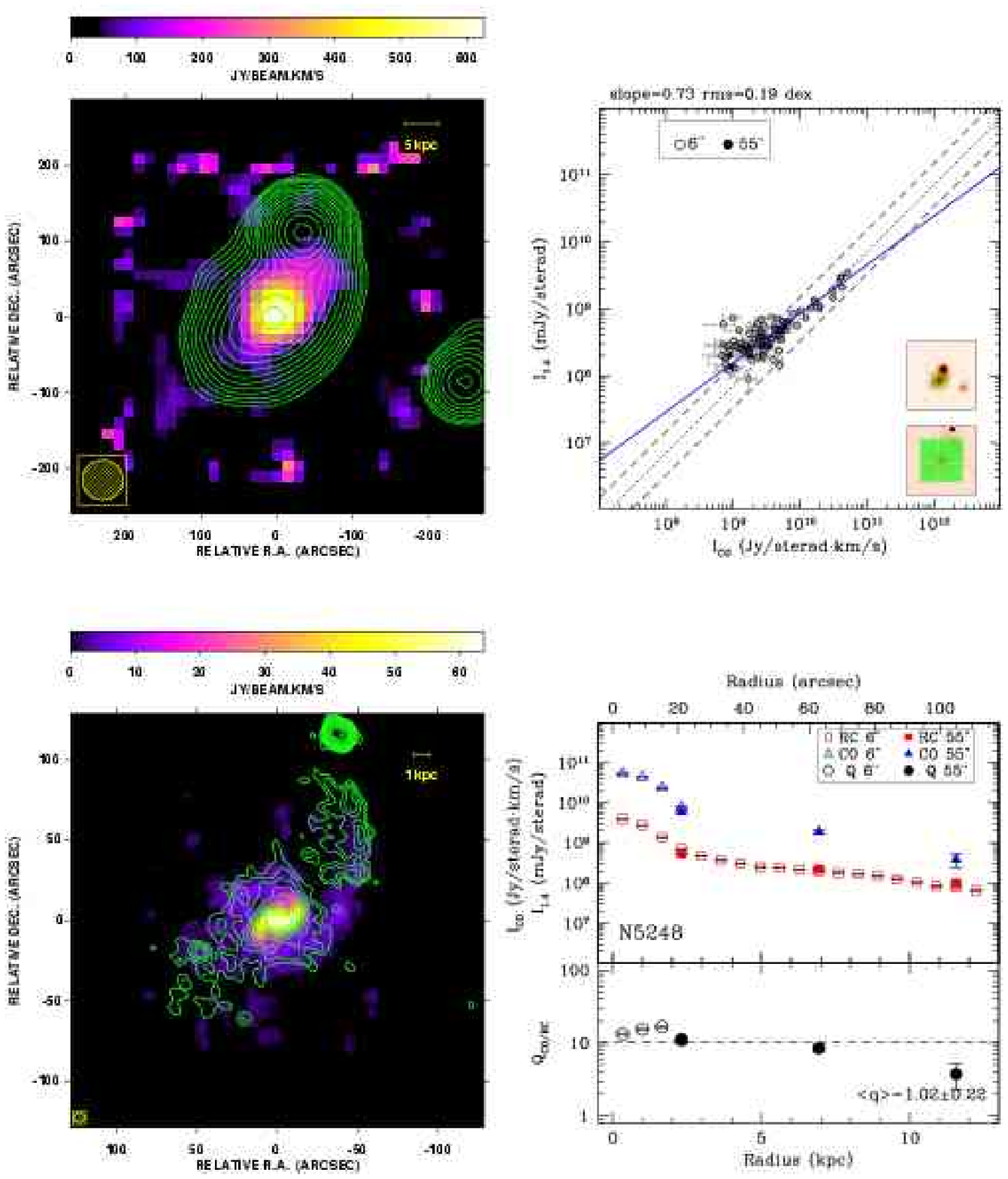}
\end{center}
\caption[] {{\bf (m) \object{NGC 5248}.} See Fig. \ref{gal1}$a$ for technical details.
}
\end{figure*}

\setcounter{figure}{1}
\begin{figure*}[]
\begin{center}
\includegraphics[width=16.5cm]{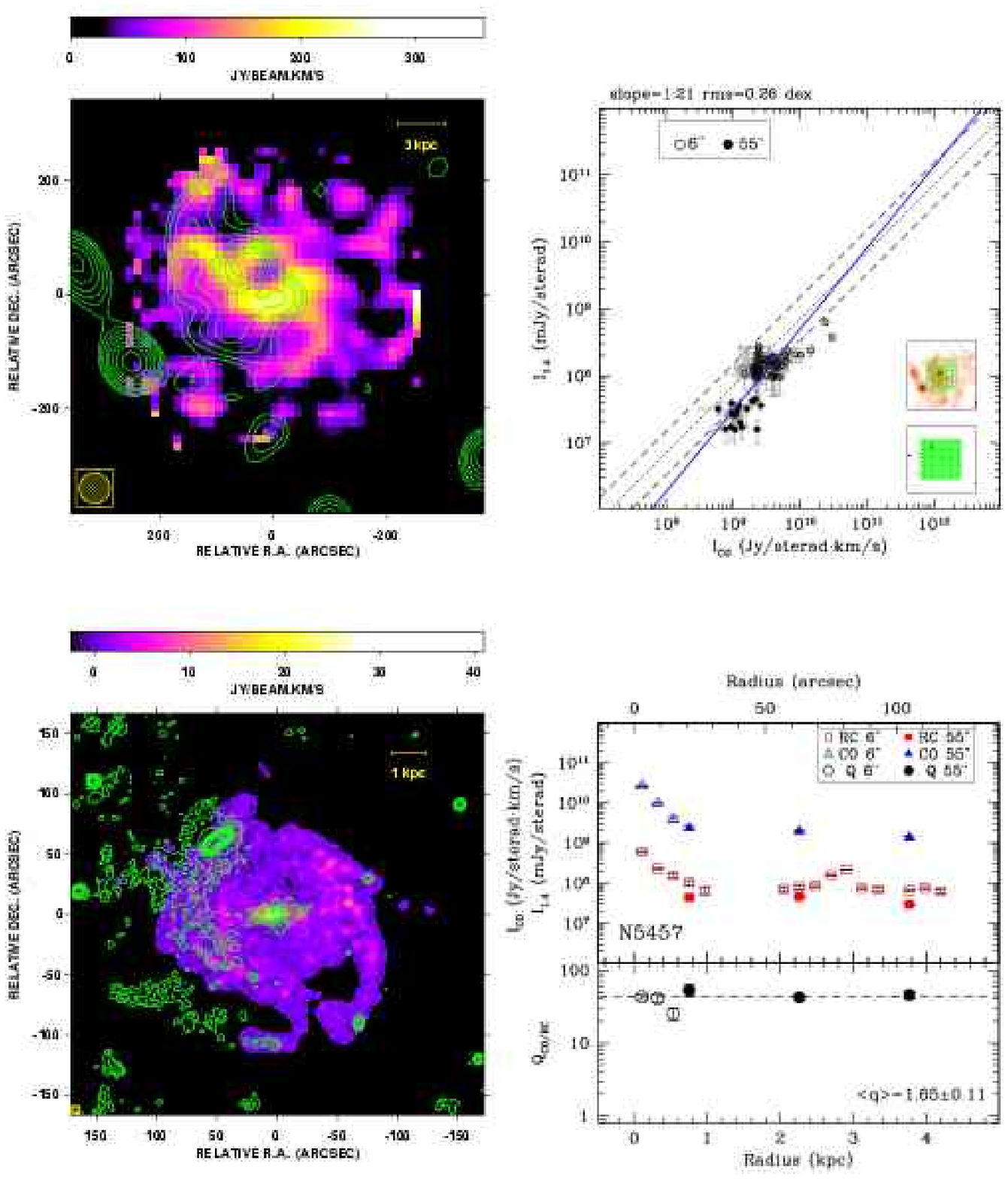}
\end{center}
\caption[] {{\bf (n) \object{NGC 5457}.} See Fig. \ref{gal1}$a$ for technical details.

}
\end{figure*}

\setcounter{figure}{1}
\begin{figure*}[]
\begin{center}
\includegraphics[width=16.5cm]{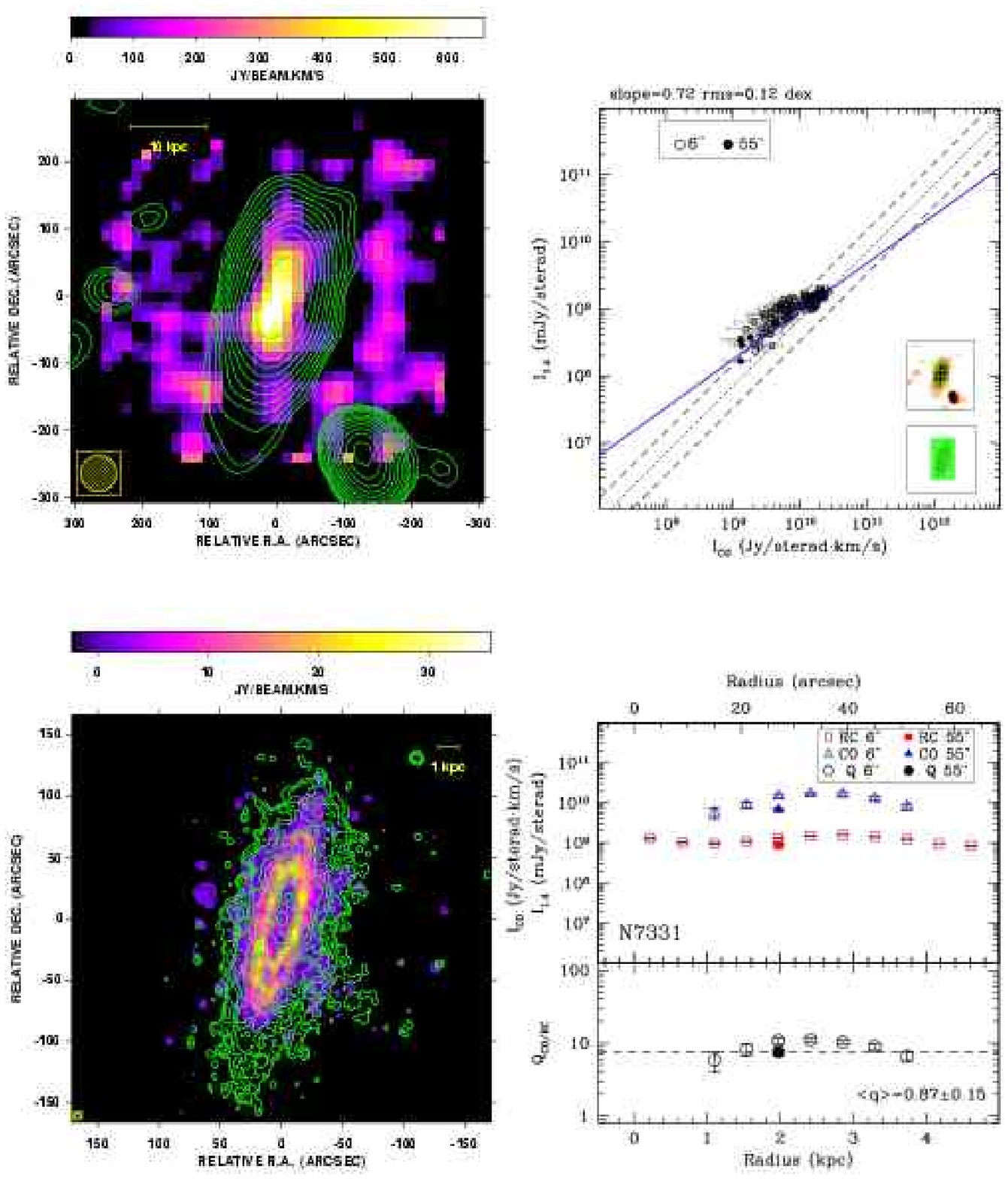}
\end{center}
\caption[] {{\bf (o) \object{NGC 7331}}. See Fig. \ref{gal1}$a$ for technical details.
}
\end{figure*}

Following the data analysis performed in Paper I, 
we compared the point-by-point RC and
 CO brightnesses across the entire galaxy disks.
We overlaid regular grids of rectangular beam-sized boxes on both the RC 
and CO images,  omitting areas with obvious background continuum 
sources. We averaged all pixel values
 within each box to calculate the 1.4 GHz RC  brightness (\i14\ ) 
and the CO integrated intensity (\ico\ ).

Figure 2 shows the results obtained for the sample of thirteen galaxies.
For each galaxy, the upper and bottom left panels show the overlay of \i14\ (contours)
 on \ico\ (color) images, respectively, at low and high resolution. 

The upper right panels plot \i14\ versus \ico\ at low (solid dots) and high (open 
dots) resolution. In the bottom right corner insets, the upper and lower panels 
show the box grids for the low and high resolution images, respectively. 
Only points above 2$\sigma$ have been plotted. 

The bottom-right panel shows the radial profiles of 
I$_{1.4}$ and I$_{\rm CO}$ (top) and their ratio Q$_{\rm CO/RC} \equiv$
~I$_{\rm CO}$(Jy sr$^{-1}$ km s$^{-1}$)/ I$_{\rm 1.4}$
(mJy sr$^{-1}$ ) (bottom). The  value of the mean and dispersion of 
q$_{\rm CO/RC}\equiv \log( Q_{\rm CO/RC})$ is given on the bottom 
right plot of the radial variation in  Q$_{\rm CO/RC}$.
For further details on the analysis technique see \matteo.

\subsection{Comments on individual galaxies.}
\label{note}
{\bf \object{NGC 0628}.}
This galaxy, seen almost face-on, has a clear 
grand-design spiral structure. 
The CO emission is strongly confined to the spiral arms \citep{regan01}.
Our RC high-resolution image (Fig. \ref{figura1a}) is  
deficient in radio emission,
 except for a bright spot corresponding to an HII region 
\citep{kennicutt80}
and an X-ray source \citep{colbert04}.
The low-resolution RC image shows emission extended over 
about 6\arcmin.
The point-by-point correlation for this source is flat  and the radial 
profile shows a decrease of the \qco\ ratio. 

{\bf \object{NGC 1068}.}
This is a nearby and luminous Seyfert 2 galaxy,  which has been extensively 
studied from radio wavelengths to X-rays.  The nuclear emission is 
clearly dominated by the radio source related to the AGN. 
 For the study of the point-by-point correlation 
we do not use the low resolution data because the source  is unresolved.
We show only the point-by-point correlation and the radial profiles
 measured in the 6\arcsec ~resolution images.
Recently, \cite{spinoglio05} found 
that the infrared emission is due to both an AGN component and to the 
starburst component in the circum-nuclear ring of $\sim$~3 kpc in size.  
In the BIMA image the CO strongest emission is 
not associated with the nucleus of the galaxy but has a ring-like 
distribution around it.
We omitted values from the fit of the 
correlation \i14\ and \ico\ measured in a central region of 
$\sim$\,15\arcsec\, where the AGN emissions of the 
compact core and prominent lobes dominate \citep{ho01}.
The resulting correlation is steeper  than the mean of all galaxies.

{\bf \object{NGC 2903}.}
This large starburst galaxy has a large-scale bar visible in 
CO and in high-resolution RC images.
The H$\alpha$ images \citep{knapen02} show star-formation along the
bar and in the
arms, mostly in the southern arm, and in the circum-nuclear region.
The bar has been studied in detail by \cite{shet02}, who
have compared the distribution of the molecular gas, traced using CO emission,
 and star formation activity, traced using the H$\alpha$ emission
line. They have found that the CO emission is brightest along the leading
edges of the bar, while the majority of the H$\alpha$ emission is offset,
with values between 200-800 pc towards the leading side of the CO.
In the point-by-point RC-CO correlation (upper-right panel of Fig. 2 (c)) 
it is possible to distinguish between data points of the central 30\arcsec, 
which show a steeper correlation,  and  those of  
 the bar emission, with a flatter correlation.
The same trend is not evident in the radial plot (lower-left panel)
where the ratio is averaged over annuli in which the 
bar-like structure is invisible.

{\bf \object{NGC 3351}.}
This is a typical 
ringed galaxy. \cite{conselice00} suggest that the nuclear starburst in this 
galaxy is due to the dynamical influence of the prominent bar.
In the BIMA image the gas appears concentrated nearly entirely 
along the ``twin peaked'' area of the nuclear region \citep{kenney92}.
Because of its small size ($<$ 1\arcmin), the galaxy is not completely 
resolved   in the low resolution image.
 The few low-resolution 
data are aligned with the extrapolation of the high-resolution data to lower 
intensity.

{\bf \object{NGC 3521}.}
This highly tilted Sbc galaxy shows strong and broad H$\alpha$
absorption around the nucleus. This absorption component is most likely
caused by a stellar absorption line of the central bulge component, 
characteristic of A-type stars.  In order
to have such a bright A-type star cluster, a large-scale starburst may
have occurred during the last $10^9$~yr \citep{sofue98}.
The high-resolution CO image shows a pronounced lack of emission 
at the center.
Our high-resolution radio image (Fig. \ref{figura1a}) shows 
the spiral galaxy with its flocculent structure, 
 and a  strong background radio source to the southwest.
The presence of this strong background 
 source degrades  the dynamic range of the image  and prevents
 us from reaching a better signal-to-noise ratio in the image. 

{\bf \object{NGC 3627}.}
This is an interacting spiral galaxy with a weak bar. 
A warp is present in the optical image and is probably 
caused by tidal interaction with the other member of the Leo Triplet (NGC 3628).
X-ray observations with Beppo SAX and Chandra suggest that the bulk
of X-ray emission has a star-forming origin \citep{georgantopoulos02}.
The molecular gas is observed in a central concentration  along  the leading 
edges of the bar, at the end of the bar, and extending over 
nearly 4\arcmin~ in spiral arms that extend from
 the ends of the bar \citep{regan01}. The same unusual 
distribution is present in the RC emission at high resolution
 (see Fig. \ref{figura1a}). 
The \qco\ radial profile (see lower right panel of Fig. 2f) shows the presence 
of a {\it{hook}} in the central region, corresponding to 
the nucleus,  which exhibits a composite LINER/H II spectrum.
\cite{filho00} reported the presence of a variable, flat-spectrum 
radio source.

{\bf \object{NGC 3938}.}
For this galaxy the RC 
contours in the bottom-right panel of Fig. 2 (g)  start from 3\,$\sigma$, since the RC emission 
at high resolution is extremely weak, with steps of a factor of $\sqrt{2}$.
It is a nearly face-on late-type spiral galaxy with well defined arms. 
It produces very faint radio emission, mostly distributed 
along the spiral arms (Fig.\ref{figura1a}).
The CO emission is also faint, so studying the correlation is 
hard, given the low dynamic range of both emissions, nevertheless 
the mean value is not far from the mean of the overall sample.

{\bf \object{NGC 4303}.}
This is a nearby double-barred galaxy, which harbors a low-luminosity 
AGN. The molecular gas properties in the central kiloparsec 
have been studied by \cite{schinnerer02}, who have found that the bulk of 
the molecular line emission comes from two straight gas lanes that run 
north-south along the leading sides of the large-scale primary bar. 
Because of the low signal-to-noise ratio of our high resolution 
RC image, obtained from archive data, we can see an analoguous  
distribution of the radio emission. 
The RC-CO point-by-point correlation is linear, and the 
radial plot of \qco\ shows the presence of the central {\it hook}, corresponding 
to the nuclear region, classified as a Seyfert 2/LINER .

{\bf \object{NGC 4321}.}
This is a well-studied, grand-design spiral galaxy, located within the Virgo
Cluster. A barred radio structure is present in our radio image 
(Fig.\ref{figura1a}).
The same distribution is visible in the high-resolution CO image.
This is evident also from the point-by-point correlation and from the radial plot.

{\bf \object{NGC 4826}.}
The 'Evil Eye' galaxy is a relatively isolated Sab spiral with a
conspicuous dust lane. \cite{ho97}, analysing spectroscopic parameters,
have classified the
nucleus as a transition object, which 
has a radio counterpart in our 20 cm image (Fig.\ref{figura1a}).
Archival observations at 
higher resolution, performed with VLA-A array, revealed the presence of 
clumps of emissions.
In the upper- and lower-right panels of Fig. 2 (l) we can see that the 
point-by-point and also 
the radial profile show a good correlation between the RC and CO emission.

{\bf \object{NGC 5248}.}
This is a nearby barred galaxy with a spectacular grand-design spiral
structure. Our radio image shows  the 
presence of a barred structure (Fig. \ref{figura1a}).
The properties of the circum-nuclear region and the relation to the host
galaxy have been discussed by \cite{knapen02}. Massive star
formation, as traced by H$\alpha$ emission, is very strong in the
circum-nuclear
region and in the (non-active) nucleus, and is located mainly along the
spiral arms and in the bar. 
The RC-CO correlation seems to lie in the mean of the sample, and 
the radial profile does not show any significant trends.

{\bf \object{NGC 5457}.}
This barred spiral galaxy is nearly face-on and tidally interacting with 
NGC\, 5474, NGC\, 5477 and Holmberg IV, which results in a distortion of the HI
distribution in its outer regions \citep{davies80}. 
In our map (see Fig. \ref{figura1a}) it presents a lopsided RC emission.
 Due to its large size ($\sim$ 10\arcmin), we could 
 infer a loss of flux from extended regions ($>$3\arcmin)
 using the B-array, but the same asymmetric 
distribution of radio emission 
is evident in the D-array image (see the top right panel of Fig. 2 (n)).
\cite{kuntz03} have studied the diffuse X-ray emission, which traces
the spiral arms and is roughly correlated with the H$\alpha$ and far-UV
emission. The X-ray source population has been studied with ROSAT HRI
and PSPC observations by \cite{wang99}, who have suggested that the
detected sources result from recent massive star formation in the outer spiral
arms.
For this unusual source the point-by-point correlation 
is flatter than the mean of the sample.

{\bf \object{NGC 7331}.}
This is a relatively highly inclined ($i~=~77^\circ$) spiral galaxy with
a ring visible in CO \citep[e.g.][]{regan01}, radio continuum
(Cowan et al. 1994), submillimeter \citep{bianchi98} and
mid-infrared images \citep{smith98}.  
Our high-resolution image clearly shows  the ring-like structure 
and the central emission due to a very low-luminosity 
radio core \citep{cowan94}, which make this galaxy 
a prototype of a submillijansky AGN \citep{filho04}. 
The RC-CO correlation has a very low dispersion and 
the \qco~ is almost constant within the ring  structure, 
which is about 1~kpc thick.

 \section{Results}

We start by summarizing the results of the 
study of RC-CO correlation from kpc to sub-kpc scale for the  
13 BIMA SONG galaxies analysed in this work.
We will then consider the whole  sample composed of the nine galaxies 
from \matteo\ and these 13, in order to improve the statistical 
significance of our results.

\subsection{Point-by-point and radial correlations of the RC and CO emissions}

The point-by-point comparison of the 
\i14\ -- \ico\ correlations are shown at both low ($>$ 1 kpc)  and high 
($<$ 1 kpc) resolution   for each galaxy 
in the upper-right panels of Fig. \ref{gal1}.
The values of the slope and of the rms, obtained 
by a least squares fit to the measured values, 
are also given. 

We find that for four galaxies (namely \object{NGC 0628}, 
\object{NGC 1068}, \object{NGC 3938} and \object{NGC 5457}) the  correlation 
is far from perfect.  
In the case of \object{NGC 0628}, the low-resolution images  show that the 
RC emission is more extended than the CO one,
 at high resolution RC emission is faint and diffuse, while 
CO is tightly confined to the spiral arms. This drop of RC emission 
causes a deviation from the mean \i14 - \ico\ relation,    
resulting in a very flat slope.
\object{NGC 1068} is a very luminous Seyfert  galaxy.
Only the high-resolution correlation is studied here, 
since in low-resolution images the source is 
unresolved. 
The emission from the central region of $\sim$\,15\arcsec\ 
is dominated by the  compact core and prominent lobes 
associated with the AGN \citep{ho01}.
Omitting these points from the fit (see 
 dashed  box in Fig. \ref{gal1}b),
the remaining points have 
a larger dispersion, and the resulting correlation is steeper than 
the mean.
For \object{NGC 3938}, due to the very low dynamic range 
of both the RC and CO emission, 
all points are clustered in the \i14 - \ico\ plot. 
\object{NGC 5457} is particularly peculiar since it exhibits a 
lopsided RC emission, which is not seen in the CO high-resolution 
image. This galaxy is $\sim$ 10\arcmin\ in diameter,  so 
 D-array observations do not suffer from losses 
of flux emitted from extended regions, while 
in B-array observations emissions from regions 
more extended than ~3\arcmin\ are lost.
Nevertheless, both low and high resolution images show  
an asymmetric distribution of radio emission, probably due 
to the tidal interactions with other galaxies. 
The slope and dispersion of the point-by-point correlation are higher 
than the mean.

The remaining nine galaxies  
(see Section \ref{note}   
for individual features of the correlation of  each galaxy)  
show slopes and dispersions of the point-by-point correlation consistent with 
the mean values for the overall sample.

In the bottom-right panels  of Fig. \ref{gal1} 
the radial profiles of \i14, \ico\ and of their ratio are shown.
The mean and dispersion in q$_{\rm CO/RC}\equiv \log (\rm I_{\rm CO}/\rm I_{\rm 1.4})$ 
are given in the bottom-right  corner. 
 Omitting \object{NGC 1068} from the calculation, 
 the average \qco\ for the remaining twelve
sources is 1.20$\pm$ 0.25; 
the dispersion in \qco\ within the individual 
sources ranges from 0.08 to 0.49. 
The findings of the point-by-point correlation 
are reflected in the radial analysis:
the four {\it anomalous} galaxies show a large deviation 
from the mean behaviour  in the radial profile.
In particular, 
\object{NGC 0628} shows a radial decrease of \qco\ of about one order of magnitude 
from central to external region. In contrast, 
\object{NGC 1068} shows the opposite trend,  but this is due to the 
fact that in the radial 
analysis the central region has not been omitted.

\begin{figure*}[]
\vspace{-0.4cm}
\begin{center}
\includegraphics[width=17cm]{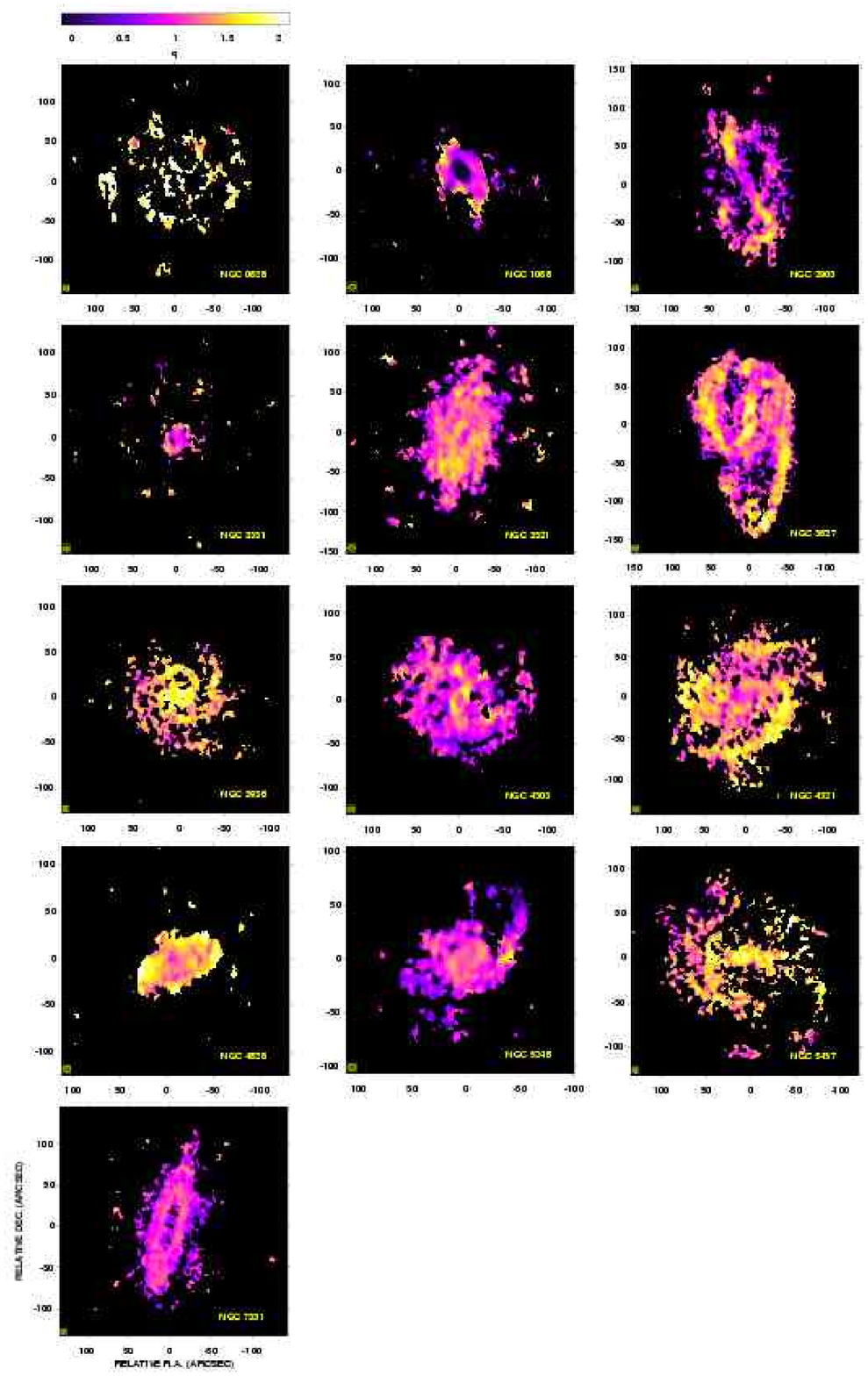}
\end{center}
\vspace{-0.5cm}
\caption[] {Images of \qco. The grey-scale is given
as a wedge at the top of the figure.  }
\label{figuraq}
\end{figure*}

Figure \ref{figuraq} shows  the images of the ratio \qco\ 
for each source. All images are shown on 
a common grey-scale, so that color variations from source 
to source indicate real variations in \qco.
The \qco\ images of \object{NGC 0628}, \object{NGC 3938} and \object{NGC 5457} clearly  show 
their deficiency of RC 
emission; \object{NGC 5457} is particularly remarkable 
 with its lopsided RC emission despite the almost symmetric CO distribution.
In the \qco\ image of \object{NGC 1068} the enhanced RC emission from 
the central AGN is evident.
There are some differences in the distribution of \qco\ among the other sources.
For \object{NGC 3351}, \object{NGC 5248} and \object{NGC 7331} there is very little variation in \qco.  
This is also evident from their low dispersions (Fig. \ref{gal1}). 
In \object{NGC 3521}, \object{NGC 4321} and \object{NGC 4826}, variations that follow the galaxies' 
spiral structure are present.
The ratios are enhanced along the dust lanes in the bars of \object{NGC 2903} and 
\object{NGC 3627}.

\subsection{Results for the overall sample}

\begin{figure*}[t]
\begin{center}
\includegraphics[height=8.5cm]{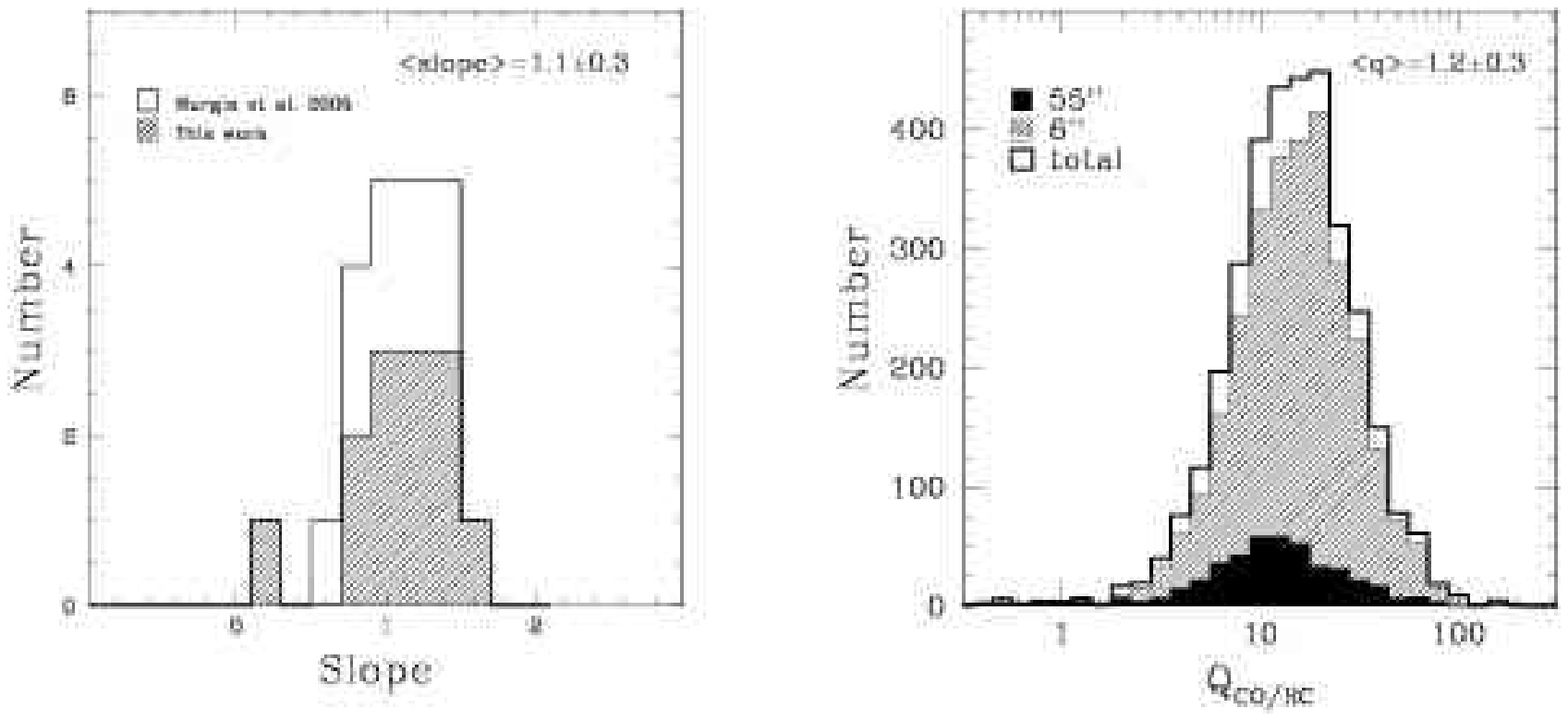}
\end{center}
\vspace{-0.5 cm}
\caption[] {{\it {Left}} Histogram of measured slopes for the overall sample of galaxies.
{\it {Right}} Histogram of all measured \qco\ grid boxes for the 22 sample galaxies. 
The means and dispersions are also given. }
\label{histoq}
\end{figure*}

In Fig. \ref{histoq} we present  histograms
 of the RC-CO correlations   
for the sample of 22 BIMA SONG galaxies. 
The left panel shows the histogram of the RC-CO correlation slopes. 
Although the slope varies from one galaxy to another (in the range of 0.29 to 1.6, 
including also two peculiar galaxies: \object{NGC 0628} and \object{NGC 1068}) the mean 
value is 1.1$\pm$ 0.3. 
In the right panel we show the histogram of \qco\ 
obtained from the  
point-by-point measures of \ico\ and \i14. 
 Averaged over all points of the grids,  $<$\qco$ >$ = 1.2 $\pm$ 0.3, 
i.e.  the uncertainty in the mean CO/RC ratio across all galaxies is less than 
a factor of 2.

\begin{figure*}[t]
\begin{center}
\includegraphics[width=12cm]{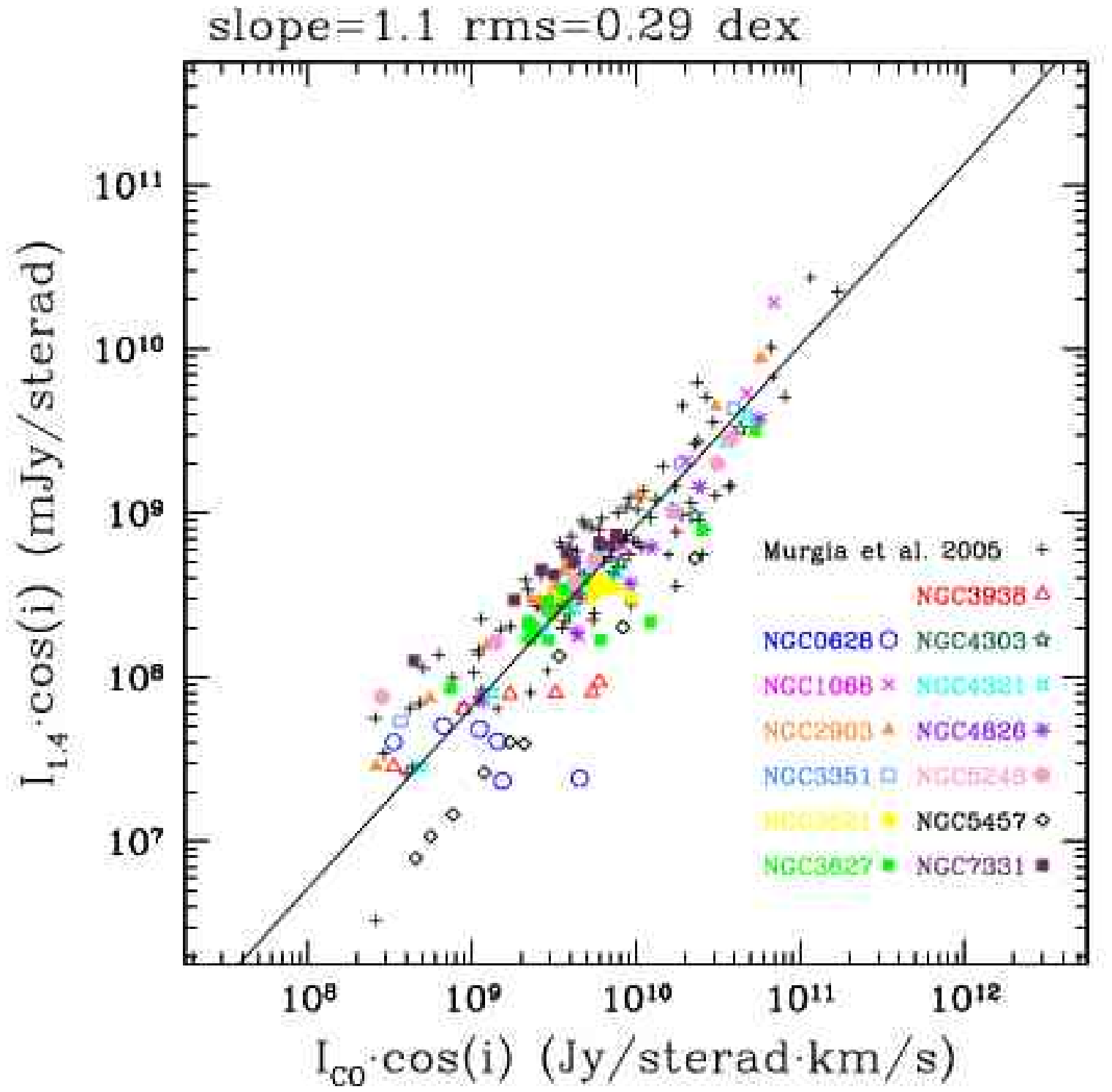}
\end{center}
\vspace{-0.5 cm}
\caption[] {Correlation between the RC and CO emission in 22 galaxies. \i14\ and  
\ico\ have been corrected for the galaxy inclinations. 
 Each point represents the average value within an
annulus. The solid line is a weighted fit to the points shown, which takes into account
the errors in both coordinates and has a slope of 1.1. The points of \object{NGC 1068} 
clearly due to the AGN radio emission are omitted from this plot. 
Different symbols represent the galaxies studied in this paper, crosses 
are the galaxies of the sample investigated in \matteo. }
\label{all}
\end{figure*}

In Fig.  \ref{all} we show the correlation between RC 
brightness and CO integrated intensity of the 22 galaxies, corrected for the inclination. 
The points correspond to values computed as annular averages, i.e. they are
 radial profiles. 
We omitted all points of \object{NGC 1068} dominated by the AGN radio 
emission from the fit. 
 The resulting correlation has a slope of 1.1 and a rms of 0.3 dex, i.e. the RC-CO 
correlation is linear with a dispersion of less than a factor of two.
The dispersion increases progressively below an RC brightness of 10$^{8}$ 
mJy/sterad. 
In this range there are three {\it{anomalous}} 
galaxies (\object{NGC 0628}, \object{NGC 3938} and \object{NGC 5457}) : 
as we have noted in previous sections,
their CO emission 
is clearly visible, while the RC emission at 1.4 GHz is so weak that the spiral structure 
is not detected at all at high resolution (see Fig. \ref{figura1a}). 
These results confirm the tightness of the CO-RC correlation  
found in \matteo, although there
 can be significant variations from source to source.

The main result of this work is that 
for the sample of 22 BIMA SONG galaxies
we find the point-by-point and radial correlations between the CO 
and RC emissions to be consistent  with the results of  \matteo.
The CO-RC correlation extends over four  orders of magnitude,
 and has a dispersion of less than a factor of 2.
This appears to be a general characteristic for spiral galaxies whose radio emission 
is dominated by starburst activity.

\section{Discussion}

\begin{figure}[t]
\begin{center}
\includegraphics[width=10cm]{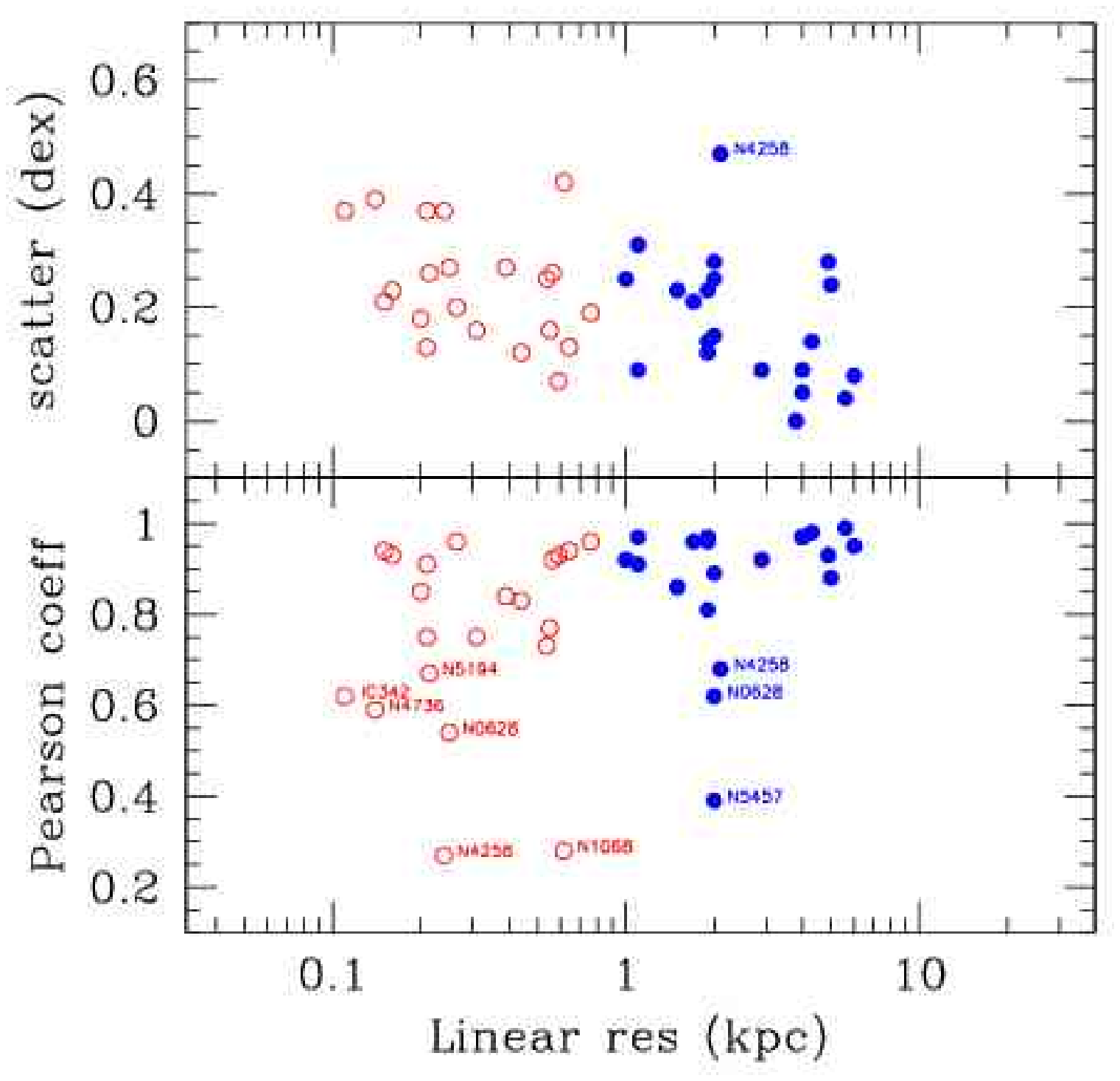}
\end{center}
\vspace{-0.5 cm}
\caption[] { Rms scatter (top panel) and Pearson coefficient
(bottom panel) of the RC/CO correlation for 21
galaxies versus linear scale of observations. Points corresponding 
to a Pearson coefficient 
smaller than 0.7 and to a scatter higher than 0.4 are labelled.} 
\label{rms_rxy}
\end{figure}
\subsection{The RC-CO decorrelation scale(s)}

One of the main aims of this work was to find out whether the RC-CO-FIR  
correlation persists at sub-kpc scales in galaxies.
The spatial resolution of our images 
 is adequate for  this  purpose. 
In fact, we can  explore the RC-CO correlation from scales of  
10 kpc down to 100 pc.
In order to estimate quantitatively the behaviour of this 
correlation we consider the rms scatter and the Pearson coefficient 
of correlations for the 22 galaxies of the  sample.
We know that the relativistic electrons diffuse from their
birth places. Therefore one expects that the spatial
correlation of the emissions should break down below
the characteristic diffusion scale-length of the radiating
electrons, L$_{\rm diff} \sim \rm v_{diff} \cdot \tau$,
 where $\rm v_{diff}$ is the diffusion velocity and $\tau$
is the electron lifetime. If we use the  
Alfv{\'e}n velocity (typically $\sim$\,10 km s$^{-1}$) in the warm ISM phase of the 
galactic disk, which may be responsible for turbulent diffusion of cosmic 
rays, as $\rm v_{\rm diff}$, we get $ \rm L_{\rm diff} \approx 1$ kpc. 
This value is extremely 
uncertain, depending on the diffusion mechanism of electrons, which is not fully 
understood and can be considered a lower limit to the scale-length (if we assume as 
the typical wind velocity in starburst 
galaxies $\rm v_{\rm w}\sim 2000 \rm 
~km ~s^{-1}$ as the diffusion speed, we obtain a much higher value of $\rm L_{\rm diff}$).

Figure \ref{rms_rxy} shows the Pearson coefficient and the scatter of 
the correlation
versus the linear scale of the observations. 
For each galaxy we plot two points, corresponding 
to the high and low resolution correlations, respectively. 
The median of the rms scatter  
increases slightly  from 0.15 in the kpc 
range  to  0.25 in the sub-kpc range,
while the median value of 
the  Pearson correlation coefficient ($r$) shows the same trend,  
 decreasing from 0.93 to 0.84.
Thus, this further quantitative estimate of 
the goodness of the correlation  
confirms the result found considering the scatter of the 
correlation, i.e. we cannot recognize any  abrupt variation
 that could indicate a possible 
 spatial scale of the decorrelation.

In Fig. \ref{rms_rxy} we have labelled points corresponding 
to a Pearson coefficient 
smaller than 0.7 and to a scatter higher than 0.4. 
We recognize those galaxies with an  {\it anomalous} 
behaviour as has been emphasized in Section 4 and in \matteo:
IC\,342, \object{NGC 0628}, \object{NGC 1068}, NGC\,4258, NGC\,4736 and \object{NGC 5457}.

The little degradation of the correlation at high resolution could  
be due to physical effects, such as the electrons' diffusion from 
their location of acceleration, or to observational effects 
(some galaxies are not completely resolved by the 
 low-resolution observations, 
and this can artificially increase the tightness of their correlation on kpc scales).

We emphasize that our study
of a statistically significant sample of 22 BIMA SONG 
galaxies does not reveal 
any irrefutable decorrelation of the RC and CO emission 
on sub-kpc scales.
This result indicates either that we have not yet probed the spatial scales 
at which the correlations break down or that there is a mechanism of 
regulation that compensates for the electrons diffusion.

\subsection{Radial behaviour}
As noted in \matteo\ the most significant deviation from the close correlation between
the FIR and RC emission is a small and generally monotonic decrease
in FIR/RC as a function of radius. \cite{marsh95}, with a study at  
a resolution of $\sim$1\arcmin, found a decrease of the FIR-RC ratio with the radial distance 
from the centers of galaxies,  which they interpreted as a smearing of the RC due to 
the propagation of energetic electrons.
The study of Marsh \& Helou has seven galaxies  in common with our sample:
 \object{NGC 0628}, \object{NGC 2903}, \object{NGC 4303},
NGC\,5055, \object{NGC 5194}, NGC\,6946 and \object{NGC 7331}. 
 These galaxies show monotonic decreases in FIR/RC 
 over 10--20 kpc in radius.  In contrast, with the sample of 
galaxies in this work we study 
the radial distribution of the \ico/\i14 ratio in 
the inner 10 kpc of the disks and we do not see any systematic 
trend with radius  (see bottom
right panels of Fig. \ref{gal1}).
\object{NGC 0628} is an exception: as we can see in Fig. \ref{gal1}~(a)
there is a radial decrease of \qco\ analogous to that observed 
in the FIR-RC ratio by \cite{marsh95}.

For the remaining galaxies,
 we have seen that the most coherent 
structures in \qco\ are found along the spiral arms, with an enhanced
\qco\ along the spiral arms and lower values in the interarm regions.

Recent {\it Spitzer} FIR imaging of four nearby galaxies 
reported by \cite{murphy06} show that the spiral structure 
is visible in infrared/radio ratio images, which show enhanced values
of FIR emission  along the arms, with local peaks centered on H\,II regions and depressed 
ratios located in the quiescent interarm and peripheral   
regions of each galaxy.
\cite{gordon04} find very similar features in 
{\it Spitzer} images of  M~81.
These images confirm the prediction of  \matteo, that 
when averaged in
azimuthal rings, the fraction of the total area in an annulus
subtended by the spiral arms is much larger at small radii than 
at large ones.  It may be that the
measured radial falloff in the FIR/RC ratio in the IRAS and ISO 
observations is primarily a consequence of the
low resolution of the FIR observations.

\section{24$\mu$m-RC high-resolution correlation with {\it Spitzer} data}

 With the launch of the {\it Spitzer Space Telescope} \citep{werner04}, 
it is now possible to map many large galaxies in the far-IR with 
unprecedented  spatial resolution  and sensitivity. 
The MIPS \citep[Multiband Imaging Photometer for Spitzer;][] 
{rieke04} observations 
at 24 $\mu$m, 70 $\mu$m and 160 $\mu$m, have PSFs 
with angular resolutions of  
$\sim$5.7\arcsec, $\sim$16\arcsec and $\sim$38\arcsec, respectively.

\begin{figure*}[t]
\begin{center}
\includegraphics[height=16cm]{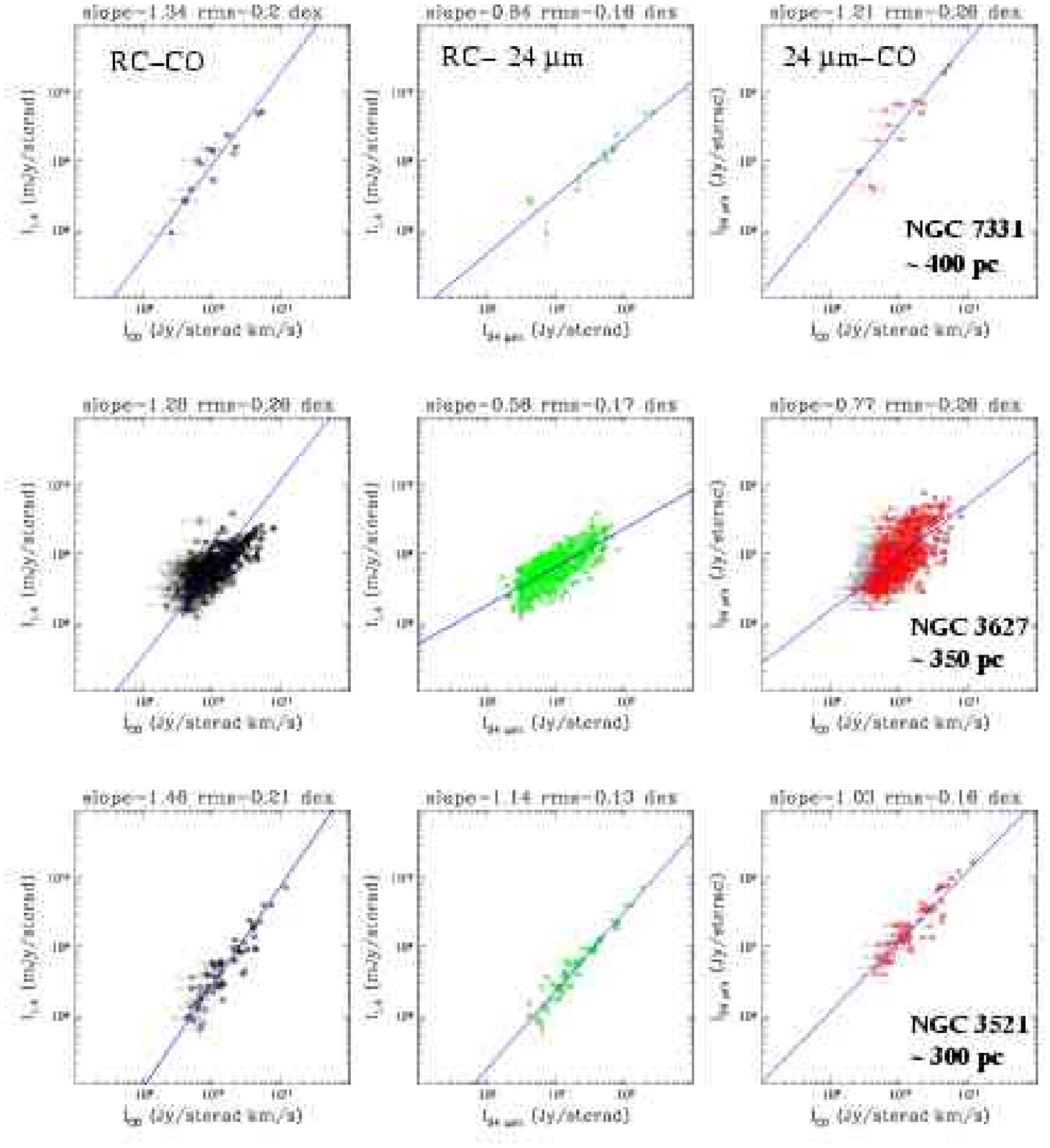}
\end{center}
\vspace{-0.5 cm}
\caption[] {RC-CO, RC-24$\mu$m and 24$\mu$m-CO correlations
 of \object{NGC 7331}, \object{NGC 3627} and \object{NGC 3521} at 6\arcsec\ resolution.
The linear resolution achieved for each galaxy is indicated in the lower-right of 
the right-most panels. 
}
\label{firrcco1}
\end{figure*}

\begin{figure*}[t]
\begin{center}
\includegraphics[height=16cm]{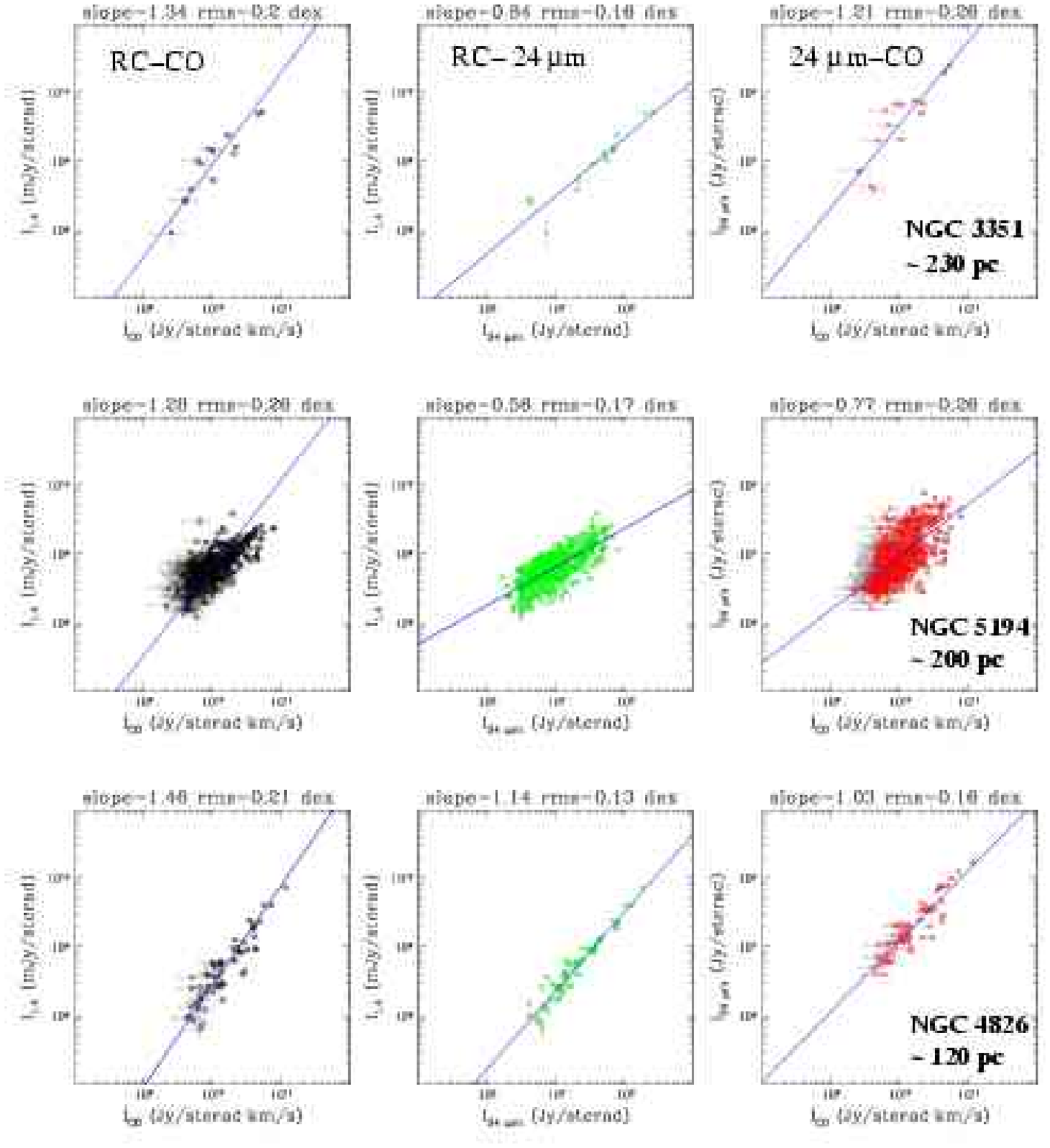}
\end{center}
\vspace{-0.5 cm}
\caption[] {RC-CO, RC-24$\mu$m and 24$\mu$m-CO correlations 
of \object{NGC 3351}, \object{NGC 5194} and 
\object{NGC 4826} at 6\arcsec\ resolution.
}
\label{firrcco2}
\end{figure*}

In this section we focus on six galaxies in our sample 
for which 
{\it Spitzer} images have been recently released:
\object{NGC 3351}, \object{NGC 3521}, \object{NGC 3627},  
\object{NGC 4826}, \object{NGC 5194} and \object{NGC 7331}.  
With 6\arcsec\ angular resolution 
we reach  spatial scales 
ranging from $\sim$ 400 pc to $\sim$ 120 pc.

\begin{figure}[t]
\begin{center}
\includegraphics[width=10cm]{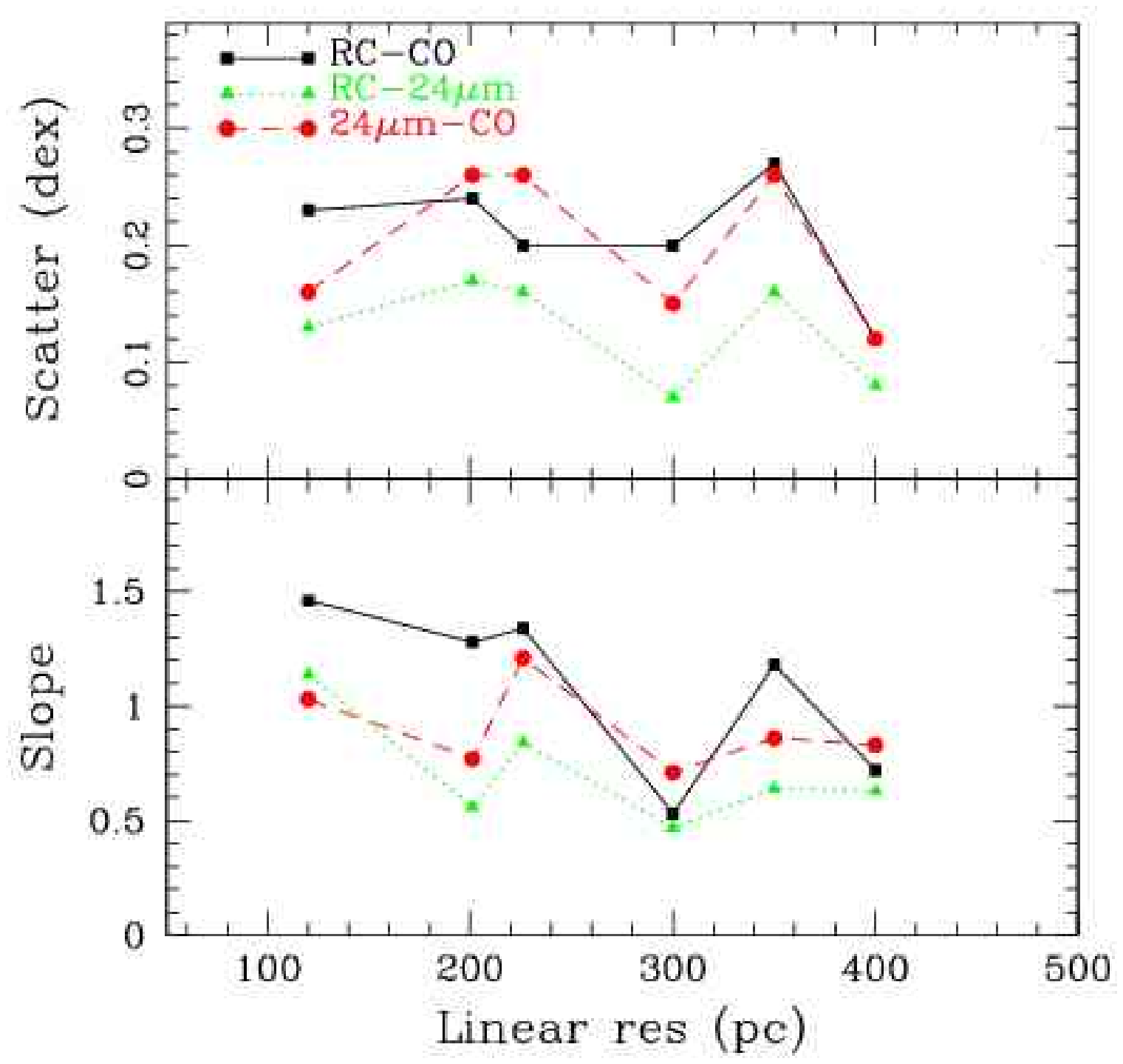}
\end{center}
\vspace{-0.5 cm}
\caption[] { Rms scatter (top panel) and slope 
(bottom panel) of RC-CO, RC-24$\mu$m and 24$\mu$m-CO correlations
versus linear resolution for six
galaxies of our sample for which {\it Spitzer} images 
have been released. } 
\label{slope_scatter}
\end{figure}

In Fig.s \ref{firrcco1} and \ref{firrcco2} 
we show the point-by-point correlations between the  RC, CO and mid-IR 
24 $\mu$m emissions for these galaxies, sorted by increasing linear resolution. 
From these plots we cannot 
recognize any systematic variation of 
correlations from lower to higher linear resolution.
The scatter and slope of these correlations are represented 
graphically in Fig. \ref{slope_scatter} as functions of 
the spatial resolution.
There is no significant variation of the slope and the scatter of  
the correlations at this spatial resolution. 
All three correlations are  comparably  
tight: they extend over two orders of magnitude
and have a scatter of less than a factor of two.
What is clearly seen is that the 24$\mu$m-RC correlation has the smaller 
scatter and somewhat flatter slope, while the scatters of RC-CO and 24$\mu$m-CO 
correlations are comparable (see Table \ref{tab5}).

\begin{table*}

\object{NGC 0628}\begin{tabular}{lccccccc}
\noalign{\smallskip}
\hline
\noalign{\smallskip}
\noalign{\smallskip}
Galaxy & Res &Slope & Scatter& Slope & Scatter& Slope & Scatter\\
   &(pc) & \multicolumn{2}c{RC-CO}   & \multicolumn{2}c{RC-24$\mu$m}  & 
\multicolumn{2}c{24$\mu$m-CO} \\
      \hline
\object{NGC 7331} &400& 0.72&0.12& 0.63&0.08&0.83&0.12\\
\object{NGC 3627} &350& 1.18&0.27& 0.64&0.16&0.86&0.26\\
\object{NGC 3521} &300& 0.53&0.1& 0.47&0.07&0.71&0.15\\
\object{NGC 3351} &230& 1.34&0.2& 0.84&0.16&1.21&0.26\\
\object{NGC 5194} &200& 1.28&0.26& 0.56&0.17&0.77&0.26\\
\object{NGC 4826} &120& 1.46&0.21& 1.14&0.13&1.03&0.16\\
Average & & 1.1$\pm$0.2&0.19$\pm$0.02 & 0.7$\pm$0.1&0.13$\pm$0.01 &0.9$\pm$0.1&0.20$\pm$0.02\\
\hline
\noalign{\smallskip}
\label{lr}
\end{tabular}
\caption{Parameters of the RC-CO, RC-24$\mu$m and 24$\mu$m-CO correlations 
at 6\arcsec. The scatter is expressed in dex.
}
\label{tab5}
\end{table*}
\hspace{0cm}

In conclusion, it seems that 
the 24$\mu$m-RC-CO correlations persist at spatial scales 
of the order of $\sim$ 100 pc.
The observed small scatter in galaxies for which 
we obtain the higher spatial resolution indicates 
that we have not yet probed the spatial scales at which 
the correlations break down. 
An increase of the scatter is observed for the galaxy 
\object{NGC 3627}, which  is characterized by a complex morphology.
Probably the presence of bar and spiral arms increases  
the scatter of the high-resolution correlations.

\section{\object{NGC 5194}: a test for the leaky box model}
\object{NGC 5194} has been recently observed with {\it Spitzer}.
\cite{calzetti05} found that the 24~$\mu$m luminosity appears to be 
closely correlated to  sites of active star formation;  
an initial look at the 24~$\mu$m-RC correlation has been 
presented by \cite{murphy06}.
We use these new FIR high-resolution images to 
test the predictions of the model proposed in \matteo.
We have proposed  
that the CO-RC-FIR correlations can be explained as a consequence
 of regulation by hydrostatic pressure.
In high-pressure regions, a higher fraction of 
the gas becomes molecular, which enhances the CO emission.
 In these regions of strong FIR emission 
the higher pressure is also manifested in a stronger magnetic field, 
which enhances the synchrotron emission. 

In this model, the ratio between
the confinement time, $t_{c}$, and the radiative lifetime, $t_{\rm syn}$,
of  the  synchrotron electrons is related to the ratio between RC and FIR emissions 
(see Eq. 8 in \matteo). In particular, 
we expect that an increase of the RC/FIR ratio will result 
in a steepening of the synchrotron spectrum.

\begin{figure}[t]
\begin{center}
\includegraphics[width=9cm]{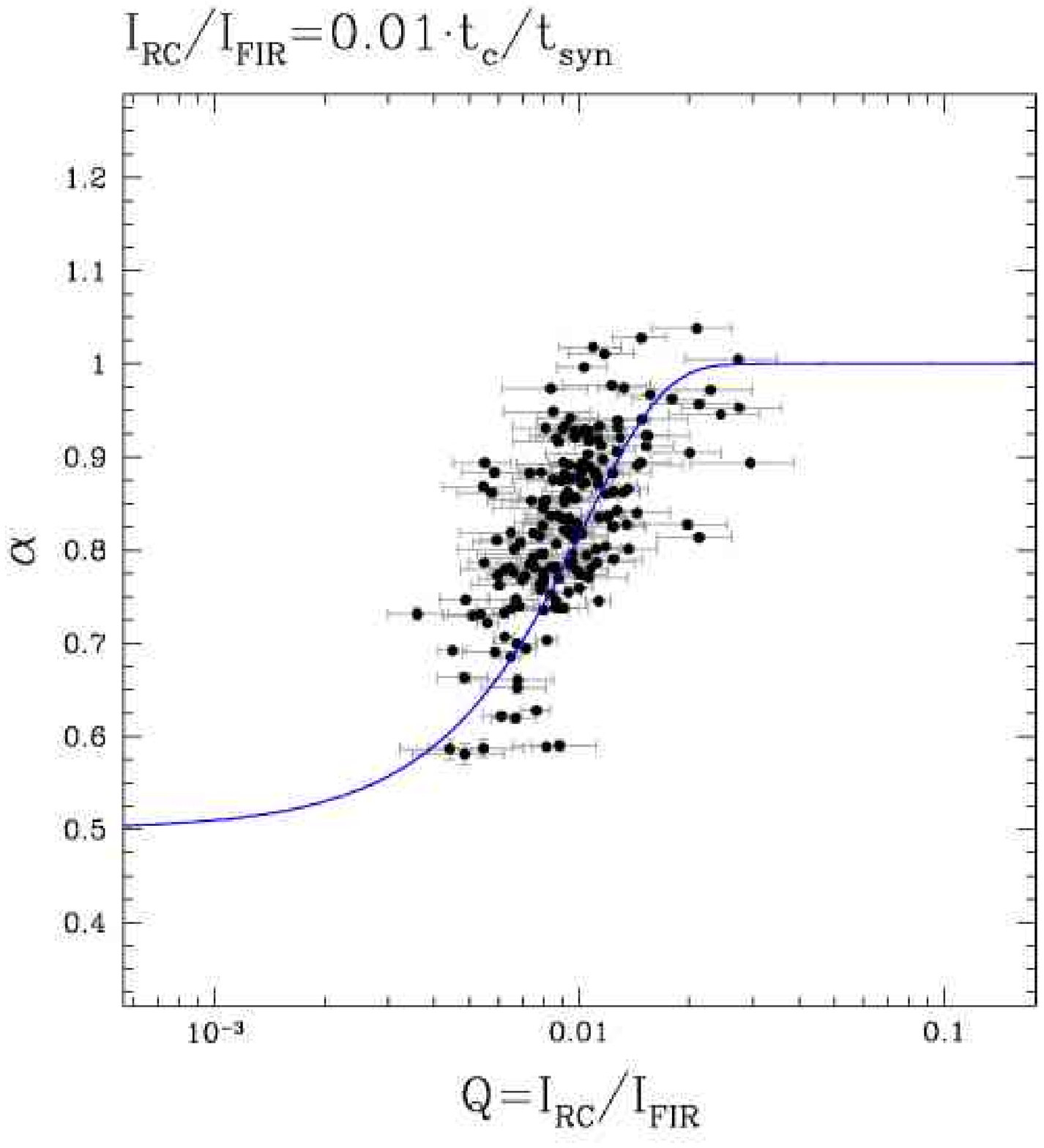}
\end{center}
\vspace{-0.5 cm}
\caption[] {Point-by-point comparison between the spectral index 
and RC-FIR ratio, measured at a resolution of $\sim$\,16\arcsec. Points 
above 3 $\sigma$ have been plotted. The line represents the fit of the 
leaky box model (see text).
 } 
\label{prova}
\end{figure}

Using the 70 $\mu$m {\it Spitzer} image of \object{NGC 5194} 
it is now possible to study the point-by-point relation between 
the spectral index and the RC-FIR ratio.
We measured the radio continuum spectral index between 1.4 and 
4.9 GHz  using the images by \cite{klein84}.
These images have 
 a resolution of $\sim$\,16\arcsec, comparable 
to that of the  70-$\mu$m observations, and incorporate short-spacing single-dish 
observations.
The relation between the radio continuum spectral index $\alpha$,
defined as I$_{\nu} \sim \nu^{- \alpha}$, and the RC/FIR ratio is 
shown in Fig. \ref{prova}. 
As the RC/FIR  ratio increases by an order of magnitude, 
a monotonic steepening of the spectral index is observed.
This behaviour  is  
expected assuming a leaky-box model of cosmic rays 
in which the injection rate of the relativistic electrons, $A$, is proportional to 
the FIR and inversely proportional to the electron confinement time.
The line shown in Fig. \ref{prova} represents the fit of the 
spectral index trend expected for a leaky box model  
with an injection spectral index of $\alpha_0$=0.5.

The exact spectral shape has 
to be computed numerically. However,  
we have two asymptotic power law regimes:
for $t_{c}\ll t_{\rm syn}$
the synchrotron emissivity is:
\begin{equation}
j_{\nu}= f(\alpha_0) \frac{C_f C_{\nu}^{\alpha_0}}{2} N_0 \cdot B^{1+\alpha_0} \nu^{-\alpha_0} 
\end{equation}

\noindent
while for $t_{c}\gg t_{\rm syn}$ the spectrum is a power law with a spectral index 
$\alpha=\alpha_0+0.5$.

 The constants $C_f$ and $C_{\nu}$ 
 depend only on fundamental physical constants:
 $C_f = 2.3 \times 10^{-22}$ erg Gauss$^{-1}$, 
$C_{\nu} = 6.3 \times 10^{18}$ erg$^{-2}$ Gauss$^{-1}$ s$^{-1}$
\citep[see e. g.][]{basicsync}. 
 
The term $f(\alpha)$ has to be calculated numerically and 
 for $\alpha_0$=0.5 is $f(\alpha)\simeq$ 3, while $N_{0}$=A$\cdot t_c$.
Here $N_{0}$ is the normalization
of a cosmic-ray electron energy spectrum $N(\epsilon) = N_0 \epsilon^{-\delta}$, 
where $N(\epsilon) d\epsilon$ is the number density of cosmic-ray electrons 
with energies between $\epsilon$ and 
$\epsilon + d\epsilon$. 
For $\alpha_0$=0.5, N$_{0}$ and $A$ are expressed in erg $\cdot$ cm$^{-3}$.
and in erg $\cdot$ cm$^{-3}$ s$^{-1}$, respectively.\\
~\\
\noindent We tentatively relate the rate of energy injection in cosmic ray electrons $A$
to the FIR emissivity, and hence to the SFR,
 introducing a proportionality constant $\eta$, so that:
\begin{equation}
A= \eta ~ j_{\rm FIR}
\end{equation}
The electron synchrotron lifetime at a given frequency is, in c.g.s. units: 
\begin{equation}
t_{syn}= 0.93 \cdot \frac{C_{\nu}^{-0.5}}{C_f} \cdot B^{-1.5} \nu^{-0.5} 
\end{equation}
Substituting  all of this into Eq. 1 we have the RC  
and  FIR emissivity related by:

\begin{equation}
\frac{j_{\nu}}{j_{\rm FIR}}=  1.4 \cdot
\eta~  C_{\nu}^{\alpha_0 -0.5} B^{\alpha_0 -0.5} \nu^{-\alpha_0 -0.5}~ \frac{\rm t_{\rm c}}{\rm t_{\rm syn}} 
\end{equation}

\noindent 

For our observations at $\nu$=1.4 GHz, assuming an injection spectral 
index of $\alpha_0$=0.5, we obtain:
\begin{equation}
\Big(\frac{I_{RC}}{I_{FIR}}\Big)_{obs}= \frac {j_{RC}}
{j_{\rm FIR}}
= 9.85 \times 10^{-10} ~\eta \frac{t_{c}}{t_{{syn}}}
\label{numer_obs}
\end{equation}

\noindent 
Expressing Eq. 2 in terms of FIR brightness, 
the value of $\eta$ allows us to infer the average electron injection rate 
at a given region of the disk:

\begin{equation}
 A= \eta \cdot \frac{4.1 \times 10^{-38}}{l_{disk}/{\rm kpc}} \frac{I_{FIR}}
{\rm MJy ~\rm sterad^{-1}} \qquad \rm erg ~\rm s^{-1} ~\rm cm^{-3}
\label{Amis}
\end{equation}

\noindent where $l_{\rm disk}$ is the thickness of the disk,
 and $I_{FIR}$ 
is the FIR brightness of the galaxy
in units of  MJy sterad$^{-1}$.

In the case of \object{NGC 5194},
from the fit shown in Fig. \ref{prova}, we  estimate the value of  
$\eta \simeq 10^7 \rm s^{-1}$. We measure an average  FIR brightness of 
25 MJy sterad$^{-1}$, which results in A$\simeq 10^{-29} \, \rm erg\,\rm s^{-1} \rm cm^{-3}$.

If the CR electrons are injected in the energy range from $\epsilon_1$ 
and $\epsilon_2$,
 this requires a injection power of W$_{el}$=$A\, \rm \ln\frac{\epsilon_2}{\epsilon_1} $, 
for $\alpha_{0}=0.5$. If we consider an electron energy spectrum ranging from $\gamma$ 1 
to $10^4$, where $\gamma$ is the electron's Lorentz factor, it results in a power 
W$_{el}\simeq 10^{-28}\, \rm erg\,\rm s^{-1} \rm cm^{-3}$.

This energetic requirement  can be compared to the SFR,   
assuming that a fraction $\xi$ of the total energy of a supernova (SN) 
 goes into CR electrons, providing a yield 
of $\xi \cdot 10^{51}$ erg per SN.
In a volume of 1~kpc$^3$ a SN rate of $\nu_{SN}$=1~SN yr$^{-1}$ yields 
\begin{equation}
W_{\rm SFR}\simeq \xi~ 1.4~ \cdot 10^{-21} \frac{\nu_{\rm SN}}{\rm yr^{-1} \rm kpc^{-3}} \qquad \rm erg ~ \rm s^{-1}~\rm cm^{-3}.
\end{equation}

\noindent \cite{condon92} calibrated the relation between the star formation and 
the SN rate:
\begin{equation}
\Big( \frac{\nu_{SN}}{\rm yr^{-1}}\Big) \approx 0.041 \Big[ \frac{SFR(M\ge 5 M_{\odot})}
{ M_{\odot} \rm yr^{-1}}\Big].
\end{equation}

\noindent Hence we obtain:
\begin{equation}
W_{\rm SFR}= \xi~ 5.7 \cdot 10^{-23}\cdot \Big[ \frac{SFR(M\ge 5 M_{\odot})}
{ M_{\odot} \rm yr^{-1} kpc^{-3}}\Big] \qquad \rm erg ~\rm s^{-1}~\rm cm^{-3}.
\label{Asfr}
\end{equation}

\noindent 
The star formation rate of \object{NGC 5194} estimated from the extinction-corrected 
H$_{\alpha}$ luminosity 
is SFR $\sim$ 4.2 $M_{\odot}~ \rm yr^{-1}$  \citep{scoville01}.
Assuming a diameter of 10 kpc and a thickness of the disk of 1 kpc 
gives $ W_{\rm SFR}= \xi \cdot ~7.2 \times 10^{-25}$ erg~ s$^{-1}$~cm$^{-3}$.
This value is  comparable to our estimate if a fraction $\xi \simeq 10^{-4}$ of 
SN energy goes into the acceleration of CR electrons.

The diffusive shock acceleration mechanism in supernova remnants is believed to 
produce the majority of Galactic cosmic rays \citep[see][ for a recent review]{hillas05}.
Nevertheless, the understanding of the electron injection is poor. 
Typically it is assumed that 10\% of the SN explosion energy goes into CR acceleration and of 
this 1\% into the electron component. Recently \cite{dyer01} studying  
TeV emission found that the current energy content in relativistic 
electrons in SN\,1006 is  about 7 $\times 10^{48}$ ergs, i.e. 10$^{-3}$ of the 
SN explosion energy. On the other hand, for Tycho's SNR 
\cite{voelk05}  found a proton injection rate of 10$^{-4}$ which, 
assuming an electron to proton ratio of $\sim$ 0.01, means an electron 
acceleration rate of 10$^{-6}$.
Thus our estimate for the conversion factor 
$\xi$ is plausible from the point of view of the energy budget 
given the observed results on individual SNRs.

\section{Summary and conclusions}

\begin{enumerate}
\item{We obtained new 1.4 GHz radio continuum images 
at 6\arcsec\ resolution for a sample of 13 nearby spiral galaxies 
belonging to the BIMA SONG survey.}
\item{We analysed the behaviour of the 
point-by-point correlation between the RC and CO 
intensities from kpc to sub-kpc scales in these objects,
 extending the previous study by \cite{matteo05} to 
22 galaxies.}
\item{For most galaxies we find that the 
RC and CO intensities are nearly linearly correlated 
(slope $\simeq$ 1.1 $\pm$ 0.3) over four orders of magnitude,  
with a scatter of a factor of 2 (0.3 dex). 

However, for four galaxies, namely \object{NGC 0628}, 
\object{NGC 1068}, \object{NGC 3938} and \object{NGC 5457}, the correlation 
is not seen. 
\object{NGC 1068} is a Seyfert galaxy whose RC emission 
is strongly contaminated by the AGN. 
\object{NGC 0628}, \object{NGC 3938} and \object{NGC 5457} 
have low CO and RC surface brightness.
The deviation from the RC-CO correlation is   
related to the fact that the RC emission is fainter than average 
at 6\arcsec\ resolution.}

\item{In the inner 10 kpc of the disks of the 22 galaxies 
analysed in this work the \qco\ ratio 
is constant with radius. The few exceptions are again \object{NGC 0628}, 
\object{NGC 1068} and \object{NGC 3938}.}
\item{In order to determine whether the goodness of the  
RC-CO correlation degrades going from 
kpc to sub-kpc scales, we studied the variation of the 
correlation scatter and the Pearson coefficient as a function 
of the spatial resolution of our images. We find that 
the median scatter increases from 0.15 to 0.25,  
going from $\sim$10 kpc to 0.1 kpc. A similar trend 
characterizes the correlation coefficient from about 0.93 to 
0.84.
We conclude that the point-by-point RC-CO correlation
deteriorates slightly when increasing the spatial resolution. However, we cannot identify 
any characteristic scale of decorrelation.}
\item{We present the point-by-point
 24~$\mu$m-RC and  24~$\mu$m-CO 
correlations at sub-kpc scales for six galaxies of our sample.
We probe the 
correlations on spatial scales ranging from $\sim$\,450 pc 
to $\sim$\,120 pc.
The RC-CO and 24$\mu$m-CO correlations 
are comparably tight,  with a scatter of $\sim$\,0.2 dex.
The 24$\mu$m-RC correlation has a scatter of only 0.1 dex 
and is somewhat flatter than the FIR-CO and RC-CO correlations.
Moreover, we find that the scatter of the correlations does not 
increase with increasing spatial resolution.
}
\item{For the galaxy \object{NGC 5194}, using the {\it Spitzer} 70 $\mu$m image,  
it has been possible to test the leaky box diffusion model proposed 
in \matteo. A point-by-point comparison between spectral index 
and RC-FIR ratio, measured at a resolution of $\sim$\,16\arcsec, shows that 
regions with a  flat spectral index have a reduced RC-FIR ratio. 
}
\end{enumerate}

\begin{acknowledgements}
We thank the referee U. Klein for useful comments which improved 
this work. The National Radio Astronomy Observatory is 
operated by Associated Universities, Inc., 
under contract to the National Science
 Fundation.
\end{acknowledgements}

\bibliography{5002}
\bibliographystyle{aa}

\end{document}